\documentclass[iop,numberedappendix,appendixfloats]{emulateapj}
\usepackage{epstopdf}
\usepackage{graphics}
\usepackage{natbib}

\newcommand{\msun}{\mbox{$\rm M_{\odot}$}}

\newcommand{\kms}{\mbox{km s$^{-1}$}}

\newcommand{\etal}[1]{{ et al.}~}
\def\kms{\ifmmode \hbox{km~s}^{-1}\else km~s$^{-1}$\fi}
\def\etal {{\it et al.}}
\def\deg      {{\ifmmode^\circ\else$^\circ$\fi} } %%% Overwrites TeX \deg

\def\h2     {H$_2$}

\def\arcsec{\hbox{$^{\prime\prime}$}}

\shorttitle{ALMA Dust Masses}
\shortauthors{Scoville \etal}

\begin{document}

\title{The Evolution of ISM Mass Probed by Dust Emission -- \\ ALMA Observations at z = 0.3 to 2}
 \author{ N. Scoville\altaffilmark{1}, H. Aussel\altaffilmark{2}, K. Sheth\altaffilmark{3}, K. S. Scott\altaffilmark{3}, D. Sanders\altaffilmark{5}, \\ R. Ivison\altaffilmark{9}, A. Pope\altaffilmark{8},  P. Capak\altaffilmark{6},
P. Vanden Bout\altaffilmark{10}, S. Manohar\altaffilmark{1},  J. Kartaltepe\altaffilmark{4},  B. Robertson\altaffilmark{11} and S. Lilly\altaffilmark{7}}

\altaffiltext{1}{California Institute of Technology, MC 249-17, 1200 East California Boulevard, Pasadena, CA 91125}
\altaffiltext{2}{AIM Unit\'e Mixte de Recherche CEA CNRS, Universit\'e Paris VII UMR n158, Paris, France}
\altaffiltext{3}{North American ALMA Science Center, National Radio Astronomy Observatory, 520 Edgemont Road, Charlottesville, VA 22901, USA}
\altaffiltext{4}{National Optical Astronomy Observatory, 950 North Cherry Avenue, Tucson, AZ 85719, USA}
\altaffiltext{5}{Institute for Astronomy, 2680 Woodlawn Dr., University of Hawaii, Honolulu, Hawaii, 96822}
\altaffiltext{6}{Spitzer Science Center, MS 314-6, California Institute of Technology, Pasadena, CA 91125}
\altaffiltext{7}{Institute for Astronomy, ETH Zurich, Wolfgang-Pauli-strasse 27, 8093 Zurich, Switzerland}
\altaffiltext{8}{Department of Astronomy, University of Massachusetts, Amherst, MA 01003}
\altaffiltext{9}{UK Astronomy Technology Centre, Science and Technology Facilities Council, Royal Observatory, Blackford Hill, Edinburgh EH9 3HJ, UK}
\altaffiltext{10}{National Radio Astronomy Observatory, 520 Edgemont Road, Charlottesville, VA 22901, USA}
\altaffiltext{11}{Department of Astronomy and Steward Observatory, University of Arizona, Tucson AZ 85721}

%\maketitle

%~~~~~~~~~~~~~~~~~~~~~~~~~~~~~~~~~~~~~~~~~~~~Accepted ApJ 2/25/13

\altaffiltext{}{}

\begin{abstract}
The use of submm dust continuum emission to probe the mass of interstellar dust and gas in galaxies
is empirically calibrated using samples of local star forming galaxies, Planck observations of the Milky Way and high redshift submm galaxies (SMGs). 
All of these objects suggest a similar calibration, strongly supporting the view that the Rayleigh-Jeans (RJ) tail of the dust emission
can be used as an accurate and very fast probe of the ISM in galaxies. We present ALMA Cycle 0 observations of the 
Band 7 (350 GHz) dust emission in 107 galaxies from z = 0.2 to 2.5. Three samples of galaxies 
with a total of 101 galaxies were stellar mass-selected from COSMOS to have $M_* \simeq10^{11}$\msun: 37 at z$\sim0.4$, 33 at z$\sim0.9$ and 31 at z$=2$.  A fourth 
sample with 6 IR luminous galaxies at z = 2 was observed for comparison with the 
purely mass-selected samples.  From the fluxes detected in the stacked images for each sample, 
we find that the ISM content has decreased a factor $\sim 6$ from $1 - 2 \times 10^{10}$\msun~ at both z = 2 and 0.9 
down to $\sim 2 \times 10^9$\msun~ at z = 0.4. The IR luminous sample at z = 2 shows a further $\sim 4$ times increase in M$_{ISM}$ compared to the 
equivalent non-IR bright sample at the same redshift. The gas mass fractions are $\sim 2\pm0.5, 12\pm3, 14\pm2 ~\rm{and} ~53\pm3$ \% for the four subsamples (z = 0.4, 0.9, 2 and IR bright galaxies). 
\end{abstract}

\medskip


\section{Introduction}

The interstellar medium (ISM) in galaxies is critical to fueling the activities of galactic star formation and AGN.
Following the peak in cosmic star formation (SF) at z = 2.5, the overall star formation rate has declined by a factor of 20 over the last 11 Gyr \citep{hop06,car13}.
 Although the ISM fuels the star formation, it is unclear if the decline
is due to: exhaustion of the galactic ISM; less efficient conversion of ISM to stars or
a reduction in galactic accretion of fresh intergalactic medium (IGM). 
At present, there exist inadequate samples of galaxies with measured 
ISM gas contents to discriminate between these possibilities. 

Over the last decade, the rotational transitions of CO  have been used to probe the molecular ISM of high redshift galaxies. 
These observations are time consuming and only $\sim$75 galaxies have been detected and imaged in CO \citep[][and references therein]{bak04,sol05,cop09,tac10,bot13,tac13,car13}. Most of the 
existing detections have been for galaxies with ISM masses exceeding $10^{10}$ \msun ~or in highly lensed systems. ALMA will  
improve this situation; still, observing times of $> 6$ hours will be required for a Milky Way-like galaxy at z $\sim$ 2.

Here, we develop an alternative approach -- using the long wavelength dust continuum to probe of ISM masses at high 
redshift \citep[as suggested by][and references therein]{sco12,eal12,mag12}. 107 galaxies in the COSMOS field at z = 0.2 to 2.5 were observed 
with ALMA in the continuum at 350 GHz ($\lambda = 850 \mu $m).  

In \S \ref{dust}), 
we present the physical and empirical basis for using long wavelength dust emission as a probe of ISM 
mass. The empirical calibration is obtained from 1) a sample of local star forming galaxies, 2) Planck observations 
of the Milky Way, and 3) high redshift submm galaxies (SMGs). All of these yield a similar rest frame 850$\mu$m luminosity per unit ISM mass with a small dispersion. 
The sample of 107 galaxies for ALMA  is presented in \S \ref{sample}. These galaxies were selected in COSMOS to have relatively constant stellar mass ($\sim 10^{11}$\msun) but span the redshift range z = 0.2 to 2.5. 26\% of the 107 galaxies were detected individually in integrations ranging from 1 to 4 minutes (with  $\sim15 -18$ telescopes
in the array and only 5 hours total time). All subsamples were significantly detected in stacked images. The observational results (\S \ref{obs} and \S \ref{app}) indicate strong evolution in the ISM contents of these galaxies from z = 2 to 0.3. 

\emph{Our results provide a strong foundation for much larger surveys of dust emission in the future 
to probe the ISM evolution at high redshift.} This approach is much faster ($\sim30\times$) than molecular line observations and more reliably calibrated 
(given the likely variations  in molecular line excitation and variations in the CO-to-H$_2$ conversion factor for higher CO transitions).

\section{Long Wavelength Dust Continuum as a Mass Tracer}\label{dust}

The FIR-submm emission from galaxies is dominated by dust re-emission of the luminosity from stars and active galactic nuclei (AGN). The luminosity at the peak of the FIR is often used to estimate the luminosity of obscured star formation or AGN. 

Equally important (but not often stressed) is the fact that the long-wavelength RJ tail of dust emission is nearly always optically 
thin, thus providing a direct probe of the total dust and hence the ISM mass -- provided the dust emissivity per unit mass  
and the dust-to-gas abundance ratio can be constrained. Fortunately, both of these prerequisites are well established from  
 observations of nearby galaxies \citep[e.g.][]{dra07b,gal11}. Theoretical understanding of the dust emissivity has  also 
 significantly improved in the last two decades \citep{dra07a}.

On the optically thin, RJ tail of the IR emission, the observed flux density is given by 
\begin{equation}
S_{\nu} \propto  \kappa_{d}(\nu)  T_{\rm d} \nu^2 {M_{\rm d}\over{d_l^2}}
\label{fnu}
\end{equation}

\noindent where $T_{\rm d}$ is the temperature of the emitting dust grains,  $\kappa_{d}(\nu)$ is the dust opacity per unit mass of dust, $M_{\rm d}$ is the total
mass of dust and $d_L$ is the source luminosity distance. (Here, we have assumed the source is at low redshift, thus not including the bandshifting and the compression of the frequency space which will be added later.) 

In nearby normal star-forming galaxies, the majority of the dust  is at $T_d \sim 20 \rightarrow 35$K,  
and even in the most vigorous starbursts like Arp 220 the FIR/submm emission is dominated by dust at temperatures $\leq 45$K. Thus variations 
in the effective dust temperature are small on a galactic scale and the observed fluxes probe the total mass of dust. 

Equation \ref{fnu} can be rewritten in terms of the specific luminosity at the fiducial wavelength $\lambda = 850\mu$m ($\nu = 353$ GHz). Let $j_{\nu}$ be the 
specific emissivity per unit volume (erg sec$^{-1}$ Hz$^{-1}$ ster$^{-1}$ cm$^{-3}$), $\kappa(\nu)$ the dust opacity per unit path length ($= n_{d} \sigma_{\nu}$) and 
vol the volume of the source. The specific luminosity is then 
\begin{eqnarray}
L_{\nu}  &=& 4\pi j_{\nu} \rm ~vol = 4\pi \kappa (\nu) B_{\nu} (T_d) \rm ~vol  \nonumber \\
 &=& 4\pi \kappa_{d}(\nu) \rho_{d} B_{\nu} (T_d) \rm vol 
\end{eqnarray}

\noindent where $\rho_{d}$ is the dust mass per unit volume in the ISM. 

For the 
RJ tail of the spectrum, this translates to
\begin{eqnarray}
\alpha_{\nu} \equiv {L_{\nu} \over{ M_{\rm ISM}}}  &=& {8\pi k \over{c^2}} \kappa_{\rm ISM}(\nu)  T_{\rm d} \nu^2   \nonumber \\
\end{eqnarray}

\noindent where $\kappa_{\rm ISM}(\nu)$ is the dust opacity per gr of ISM mass (i.e.  $\kappa_{\rm ISM}(\nu) = \kappa_{d}(\nu) \rho_{d} / \rho_{\rm ISM}$). Thus, 
\begin{eqnarray}
\alpha_{\nu} \equiv {L_{\nu} \over{ M_{\rm ISM}}} =  \kappa_{\rm ISM}( \nu)  T_{\rm d} \left({\nu \over{ \nu_{850\mu \rm m}}}\right)^2 \times  \nonumber \\
 9.54\times 10^{20} \rm erg ~sec^{-1} Hz^{-1} \msun^{-1} .
\end{eqnarray}\label{lnu}

The long wavelength dust opacity can be approximated by a power-law in wavelength 
\begin{equation}
\kappa_{\rm ISM} (\nu)  = \kappa_{\rm ISM} ( \nu_{850\mu \rm m}) (\lambda/850\mu \rm m)^{-\beta} 
\end{equation}\label{kappa}

The long wavelength opacity per unit mass (Eq. \ref{kappa}) can be investigated using submm observations of nearby galaxies and the Galactic ISM, and then applied to the high redshift galaxies. These local calibrations allow one to ascertain a best estimate for $\kappa_{\rm ISM}(\nu)$ and to determine if significant 
variations arise between atomic and molecular phases of the ISM. 
There are two aspects  to this calibration: 1) the spectral index of $\kappa_{\rm ISM}(\nu)$ is required to relate flux measures in the different wavelength bands and at different  redshifts and 2) $\kappa_{ISM}( \nu_{850\mu \rm m})$, the dust opacity per unit mass at a fiducial wavelength, here chosen to be $\lambda = 850\mu$m ($ \nu_{850\mu \rm m} \equiv 350$ GHz).

\subsection{The Dust Submm Spectral Index -- $\beta$}

The overall spectral slope of the rest frame submm dust emission flux density (Eq. \ref{fnu}) is observed to
vary as $S_{\nu} \propto \nu^{\alpha}$ with $\alpha = 3 - 4$. Two powers of $\nu$ are from the RJ dependence;  the 
remainder is due to the frequency variation in $\kappa_{\rm ISM}(\nu)$ ($\propto \nu^{\beta}$). 
Most theoretical models for the dust have opacity spectral indices of the $\beta$ = 1.5 to 2 \citep{dra11}. Empirical fits to the observed long wavelength
SEDs suggest $\beta$ = 1.5 to 2  \citep{dun01,cle10} for local galaxies. For high z 
submm galaxies, the apparent spectral index is $\alpha =$ 3.2 - 3.8 but in some cases the shorter wavelength point 
is getting close to the IR peak in the rest frame and 
therefore not strictly on the RJ tail. Probably the best determination is that of \cite{cha09} who used their $\lambda =1.1$ survey to find
$< \beta > = 1.75$ for 29 SMGs with a median z = 2.7.

Recently the Planck mission has provided a robust determination of $\beta$ with 7 bands (at $\lambda$ = 3 mm to 100 $\mu$m) observing the submm dust emission from the Galactic ISM. The \cite{pla11a} analyzed data on the Taurus 
cloud complex and derived $\beta = 1.78 \pm 0.08$, including both atomic and molecular ISM regions.  Using the very extensive Planck data throughout the Galaxy, \cite{pla11b} finds $\beta = 1.8\pm 0.1$ with no significant difference between the HI and H$_2$-dominant regions.

In the remainder of this paper, we adopt $\beta = 1.8$ as derived by the very high quality Planck observations of the Galaxy. (This $\beta$ is also 
consistent with the prior extragalactic determinations, as summarized earlier, within their uncertainties.)

\subsection{Calibration of $\kappa_{\rm ISM}( \nu_{850\mu \rm m})$ from Local Galaxies and Low-z ULIRGs}\label{lowz_galaxies}

In order to empirically calibrate the dust opacity per unit mass of ISM we make use of galaxy samples for which 
both the submm dust emission and ISM masses are well-determined globally (or where the measurements 
refer to the same areas in the galaxy). 

Our local galaxy sample includes SCUBA data for 12 galaxies -- 3 from the SINGS survey \citep{dal05} and 9 from ULIRGs survey \citep{cle10}. 
The ULIRG sample is in fact probably most relevant since these galaxies are closer in IR luminosity to the high redshift galaxies and very importantly, their 
emission is compact so we can be confident that the total 850$\mu$m flux and ISM masses encompass the same regions. (As a reliability check for the submm measurements we required observed fluxes at both 450 and 850$\mu$m and that their ratios yield a reasonable spectral index.) The submm fluxes and ISM masses for 
the low-z galaxy sample are provided in Table \ref{tab:local_gal}. 

\begin{figure}[ht]
\plotone{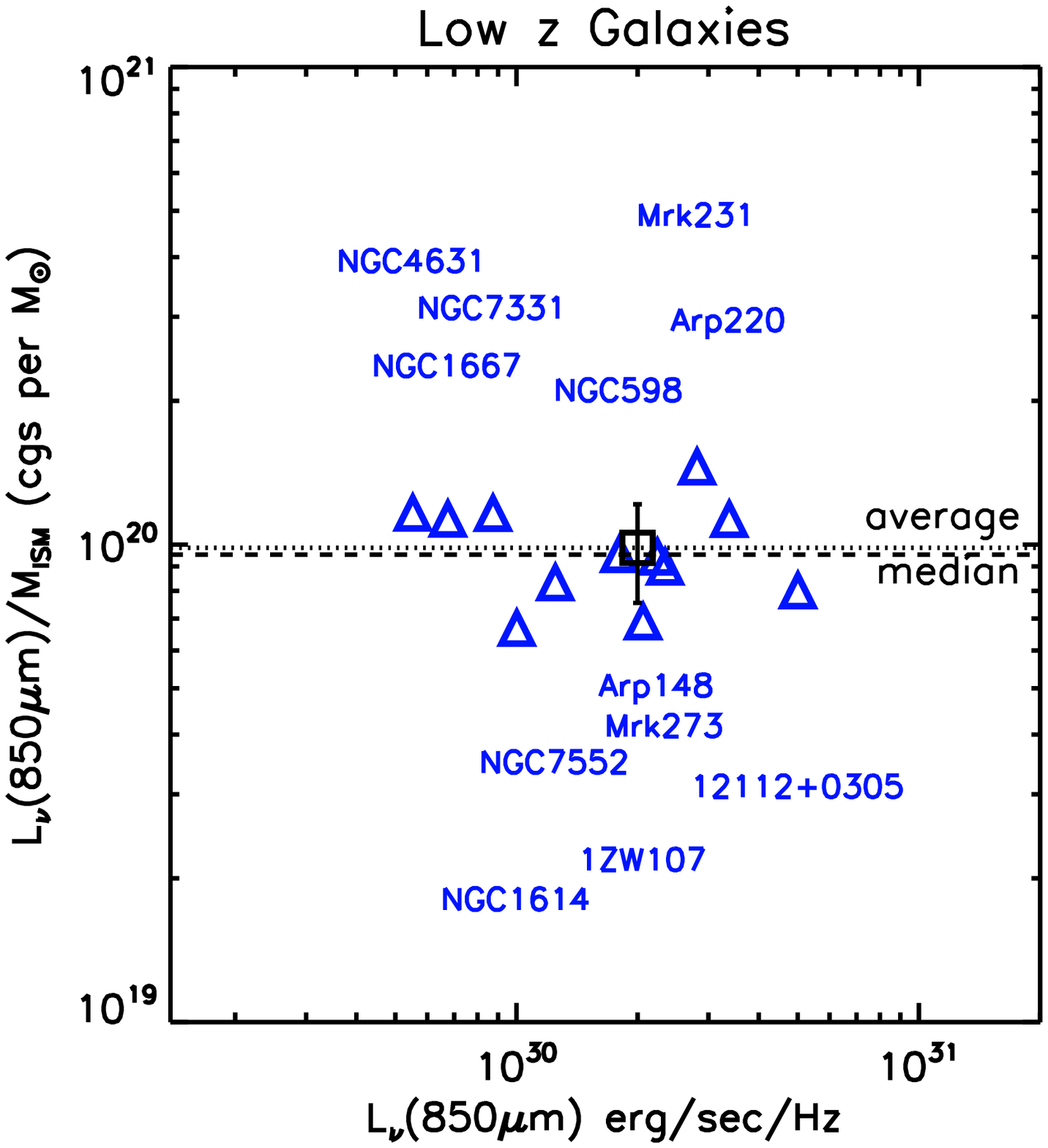}
\caption{The ratio of $ L_{\nu}$ at 850\,$\mu$m to $M_{\rm ISM}$ is shown for a
sample of low-$z$ spiral and starburst galaxies from \cite{dal05}
and \cite{cle10}. The average and median values for the
sample are shown by horizontal lines. The mean ratio shown by the error bars is  $1.00\pm0.23\times10^{20} \rm ergs~ sec^{-1} Hz^{-1} \msun^{-1}$. Typical uncertainties in the derived ISM masses are $\pm50$\% due to variation in the CO to H$_2$ conversion factor.}
\label{local_gal} 
\end{figure}

The 850$\mu$m fluxes were converted to specific luminosity $L_{\nu(850\mu \rm m)}$, 
using
\begin{eqnarray}
L_{\nu}&=&10^{-23}~4\pi d_{L}^2~S_{\nu}({\rm Jy})  \nonumber \\
& =& 1.19\times10^{27}~S_{\nu}({\rm Jy})~d_{L}^2 ({\rm Mpc}) ~\rm ergs~ sec^{-1} Hz^{-1}
\end{eqnarray}

\noindent and ratioed to the ISM masses (HI \& H$_2$) (Fig. \ref{local_gal}).
Based on the data shown in Fig. \ref{local_gal}, we find a constant of proportionality between the 850$\mu$m specific luminosity and the ISM mass

\begin{eqnarray}
  \alpha_{850\mu \rm m} &=&  {L_{\nu_{850\mu \rm m}} \over M_{\rm ISM}}  \nonumber \\
  &=& 1.0\pm0.23\times10^{20}    \rm ergs~ sec^{-1} Hz^{-1} \msun^{-1}. \label{alpha_one}
\end{eqnarray}

\begin{deluxetable}{lrrrrr}[ht]
\tablecaption{Low-$z$ galaxies with submm \& ISM data}
\tablewidth{0pt}
\tablehead{  \colhead{Galaxy} & \colhead{Distance} & \colhead{S$_{\nu}(450\,\mu$m)} & \colhead{S$_{\nu}(850\,\mu$m)} & \colhead{$\log M_{\rm HI}$} & \colhead{$\log M_{\rm H_2}$} \\
\colhead{}  & \colhead{Mpc} & \colhead{Jy} & \colhead{Jy }  &  \colhead{\msun} &  \colhead{\msun} }
\startdata
\\
NGC\,4631    &   9.0  & 30.7  & 5.73  &  9.2  &  9.5 \\
NGC\,7331    &  15.7  & 18.5  & 2.98  &  9.4  &  9.7 \\
NGC\,7552    &  22.3  & 20.6  & 2.11  &  9.7  & 10.0 \\
NGC\,598     &  76.0  &  2.3  & 0.26  &  9.8  & 10.1 \\
NGC\,1614    &  62.0  &  1.0  & 0.22  &  9.7  & 10.0 \\
NGC\,1667    &  59.0  &  1.2  & 0.16  &  9.3  &  9.6 \\
Arp\,148     & 143.0  &  0.6  & 0.09  &  9.9  & 10.2 \\
1ZW107       & 170.0  &  0.4  & 0.06  & 10.0  & 10.3 \\
Arp\,220     &  79.0  &  6.3  & 0.46  & 10.0  & 10.3 \\
12112+0305   & 293.0  &  0.5  & 0.05  & 10.3  & 10.6 \\
Mrk\,231     & 174.0  &  0.5  & 0.08  &  9.8  & 10.1 \\
Mrk\,273     & 153.0  &  0.7  & 0.08  &  9.9  & 10.2 \\
\\
\enddata
\tablecomments{Submm flux measurements and ISM masses from \cite{dal05}, \cite{cle10}
and the NASA Extragalactic Database. HI masses were set to 50\% of the molecular gas masses. Uncertainties in the 
derived H$_2$ masses are dominated by the assumed CO to H$_2$ conversion factor which is uncertain by $\sim\pm25$\% even for normal low-level SF galaxies.}\label{tab:local_gal}
\end{deluxetable}

\subsection{$\kappa_{\rm ISM} (850\mu \rm m)$ in the Galaxy from Planck}\label{pla}

The Planck measurements of the submm emission from the Galaxy provide both very high photometric 
accuracy and the ability to probe variations in the opacity to mass ratio between atomic and molecular phases, 
and with Galactic radius. The latter provides a weak probe of metallicity dependence.

In the Taurus complex, the \cite{pla11a} obtained 
best fit  ratios of $\tau_{250\mu \rm m} / N_{\rm H} = 1.1 \pm 0.2 ~\rm ~and~ 2.32 \pm 0.3 \times 10^{-25}$ cm$^{2}$ for the 
atomic and molecular\footnote{We note that the estimation of the molecular gas masses by \cite{pla11b} assumed a value X$_{CO} = 2\times10^{20}$ H$_2$ cm$^{-2}$ (K km/s)$^{-1}$ based on 
Fermi gamma ray observations. However, virial mass
estimates for the GMCs indicate a best X$_{CO} = 3.6\pm0.2\times10^{20}$ \citep{sco87}, implying higher H$_2$ masses and therefore a 
value $\tau_{250\mu \rm m} / N_{\rm H} = 1.3  \times 10^{-25}$ cm$^{2}$, i.e. more similar to the atomic phase value.}  phases, respectively. (The former is consistent with a much earlier determination of $1 \times 10^{25}$ cm$^{2}$ \citep{bou96} for the diffuse ISM.)
 
Using Planck data from the Galaxy, \cite{pla11b} found $\tau_{250\mu \rm m} / N_{\rm H} = 0.92 \pm 0.05 \times 10^{-25}$ cm$^{2}$ near the solar circle with 
no significant variation in Galactic radius. This includes the inner galaxy -- where the molecular phase dominates and the 
mean dust temperatures are higher -- in addition to the solar circle and outer radii where HI dominates. It would appear that there is no strong evidence
of variation in the opacity per unit mass with either ISM phase, dust temperature or metallicity. (The overall range of metallicity probed in the Milky Way is not large so this issue would need to be reexamined using data from low metallicity dwarf galaxies.)

In the following, we adopt $\tau_{250\mu \rm m} / N_{\rm H} \simeq 1.0 \times 10^{-25}$ cm$^{2}$.
To convert the scale factors from the Planck analysis to $\kappa_{\rm ISM}( \nu_{850\mu \rm m})$, we note that N$_{\rm H} = \Sigma_{\rm ISM} / (1.36 \rm m_H)$ where $\Sigma_{\rm ISM}$ is the total mass column density of
ISM and the factor 1.36 is to account for He at 9\% abundance. Using $\beta = 1.8$ to scale from 250 to 850 $\mu$m implies an additional factor 0.11 and thus  
\begin{eqnarray}
\kappa_{\rm ISM} ( \nu_{850\mu \rm m})&=& { \tau_{250\mu \rm m} \over{ N_{\rm H}}}  {0.11 \over{1.36\rm m_H }}  \nonumber \\
 &=& 4.84\times 10^{-3} \rm gr^{-1} cm^{2} .
\end{eqnarray}\label{kap}

The specific luminosity emitted per unit mass of ISM is then given by 
\begin{eqnarray}
  \alpha_{850\mu \rm m} =  {L_{\nu_{850\mu \rm m}} \over M_{\rm ISM}}  &=& 4 \pi \kappa_{\rm ISM}(\nu_{850\mu \rm m}) B_{\nu} (T_d) \nonumber \\
 &=& 0.79\times 10^{20}  \rm ergs/sec/Hz/\msun
\label{planck_alpha}
\end{eqnarray}

\noindent where the Planck function was evaluated with $T_d = 25$ K as discussed immediately below (\S \ref{redshift}). (Eq. \ref{planck_alpha} uses the actual Planck function, not the RJ approximation.) 

The value 
of the constant of proportionality between the 850$\mu$m flux and the ISM mass in the Galaxy found by the Planck Mission is therefore
within 20\% of that found earlier in the sample of nearby galaxies.

\subsection{Redshifted Submm Fluxes}\label{redshift}

\begin{figure}[ht]
\epsscale{1.}  
\plotone{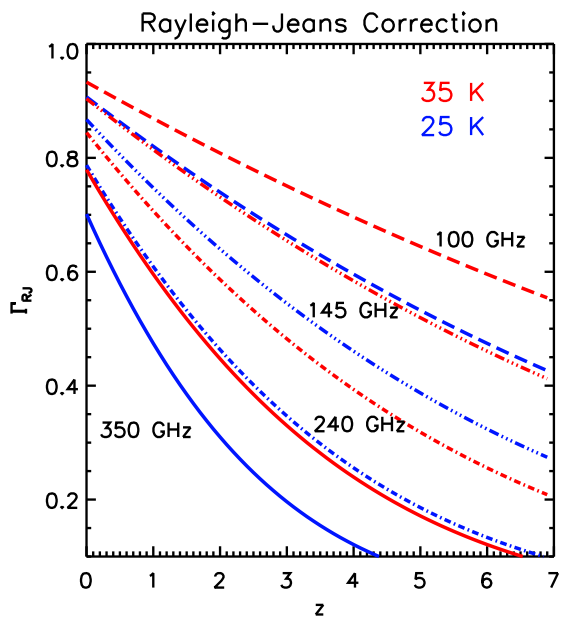}
\caption{The RJ correction factor $\Gamma_{RJ}$ is shown as a function of redshift for 4 observing frequencies and dust temperatures of 25 and 35 K. }
\label{rj} 
\end{figure}

For high redshift, one may estimate the expected submm continuum fluxes using the calibration of specific luminosity  at 850 $\mu$m to ISM mass from Equation \ref{alpha_one}
and power law with $\beta = 1.8$ for the long wavelength dust opacity. The observed flux density is 

\begin{eqnarray}
  S_{\nu_{obs}} &=& \alpha_{850\mu \rm m} ~  M_{\rm ISM}  ~ (1+z)^{4.8} ~ \left({\nu_{obs} \over{ \nu_{850\mu \rm m}}}\right)^{3.8} ~  \nonumber \\
  && \times ~{\Gamma_{RJ}  \over{\Gamma_{0}}}~  {1 \over 4 \pi d_L^2} 
  \end{eqnarray}
 
 \noindent where the additional factor of (1 + z) beyond that associated with the rest frame SED is due to the compression of observed frequency space relative to that 
in the rest frame. 

$\Gamma_{RJ}$ is the correction factor for departure from the RJ $\nu^2$ dependence as the observed emission approaches 
the SED peak in the rest frame. This can be a substantial correction; it is given by :
\begin{eqnarray}
\Gamma_{RJ}(T_d,\nu_{obs}, z) &=&  {h \nu_{obs} (1+z) / k T_d \over{e^{h \nu_{obs} (1+z) / k T_d} -1 }} 
  \end{eqnarray}

\noindent and shown in Figure \ref{rj} for dust temperatures of 25 and 35 K. [$\Gamma_0$ is the value of $\Gamma$ appropriate to 
the low z $\lambda= 850\mu \rm m$ used to calibrate $\alpha_{850\mu \rm m}$; nominally $\Gamma_0 = 0.71$ (see Fig. \ref{rj}).] 

In the following, we assume $T_d = 25$ K to characterize the bulk of the ISM mass. (In the 
Planck observations of the Milky Way the median was $T_d \simeq 18$ K \citep{pla11b} but more active galaxies are likely to have somewhat higher 
dust temperatures.) \cite{dun11} derived dust temperatures in the range 17 to 30 K for 1867 galaxies at z $< 0.5$ in the Herschel-ATLAS survey. Very similar distributions of 
dust temperatures were found by \cite{dal12} for 61 galaxies in the Herschel KINGFISH survey and \cite{aul13} for 254 galaxies in the Herschel Virgo Cluster Survey. 

For a flux density measurement at observed frequency $\nu_{obs}$, 

  \begin{eqnarray}
  S_{\nu_{obs}} &=& 0.83 ~ {M_{\rm ISM} \over 10^{10}\msun} ~ (1+z)^{4.8} ~   \left({\nu_{obs} \over{ \nu_{850\mu \rm m}}}\right)^{3.8}   \nonumber \\
 && \times ~{\Gamma_{RJ} \over{\Gamma_{0}}}~ \left({Gpc \over  {d_{L}}}\right)^{2}  ~mJy \\
\rm for  && \lambda_{rest} \gtrsim 250  ~\mu \rm m  \nonumber
  \end{eqnarray}

\noindent where we have used $\alpha_{850\mu \rm m}$ given by Eq. \ref{alpha_one}. We note that the empirical calibration of $\alpha_{850}$ was obtained from z $\simeq 0$ galaxies; however,  there is still a non-negligible 
RJ departure ($\sim 0.7$) which must be normalized out  (i.e. the y-axis intercept in Figure \ref{rj}). This is the term $\Gamma_0 = \Gamma_{RJ} (0,T_d,\nu_{850})$ in the equation above. 

The restriction $ \lambda_{rest} \gtrsim 250  ~\mu \rm m$ is intended to ensure that one is on the RJ tail and that the dust is likely to be optically thin. If the dust is extremely 
cold one might need to be more restrictive and in the case of the most extreme ULIRGs the dust is probably optically thick 
to even longer wavelengths.  Analogous expressions can easily be obtained for the other ALMA bands. 

\begin{figure}[ht]
\epsscale{1.}  
\plotone{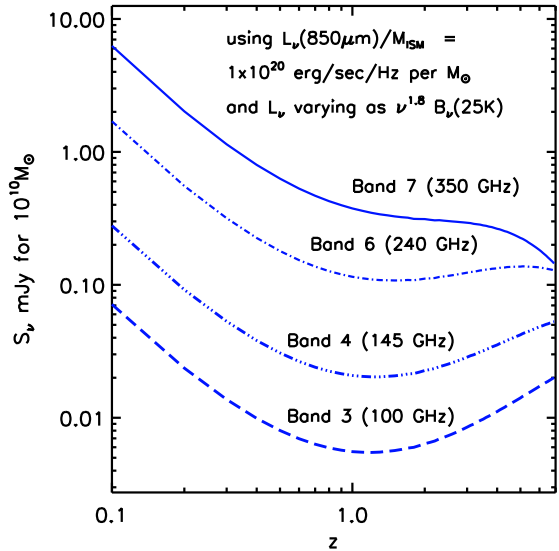}
\caption{The expected continuum fluxes at 100, 145, 240 and 350 GHz for M$_{ISM} = 10^{10}$\msun ~derived using the empirical calibration embodied in Equation \ref{alpha_one}, 
an emissivity power law index $\beta = 1.8$ and including the RJ departure coefficient $\Gamma_{RJ} (25K)$. Since the point source sensitivities of ALMA 
in the 3 bands is fairly similar, it is clear that the optimum strategy is to use Band 7 out to z $\sim 2$, before moving to lower frequency bands to avoid the RJ correction issues.}
\label{alma_obs} 
\end{figure}

Figure \ref{alma_obs} shows the expected flux as a function of redshift
for the ALMA bands at 100, 145, 240 and 350 GHz (Bands 3, 4,  6 and 7). At low z, the increasing luminosity 
distance leads to reduced flux as z increases. However, above z = 1 the well known "negative k-correction" causes the flux per unit ISM mass to increase at higher
z as one moves up the far infrared SED towards the peak at $\lambda \sim 100 \mu$m. Figure \ref{alma_obs}
shows that the 350 GHz flux density plateaus and then decreases above z = 2; this is due to the fact that at higher redshift, 350 GHz is approaching the rest frame far infrared peak (and no longer on the $\nu^2$ RJ tail). This is the factor $\Gamma_{RJ}$ coming in for 25 K dust.

 At  redshifts above 2.5, Figure \ref{alma_obs} indicates that one needs to shift to a lower frequency  
band, e.g.  240, 145 or 100 GHz, in order to avoid the large and uncertain $\Gamma_{RJ}$ corrections. Since future studies similar to that pursued here 
will push to higher redshifts, we have included the lower frequency bands in Figures \ref{rj} and \ref{alma_obs}. 

\subsection{Comparison with High z SMGs}

The empirical  value for $\alpha_{850}$ ($=  L_{\nu_{850\mu \rm m}} / M_{\rm ISM} $) obtained in \S \ref{lowz_galaxies} and \ref{pla} based on local galaxies and Planck 
observations of the Milky Way pertain to low redshift galaxies. We now compare these local calibrations with observations for a large sample of 
SMGs from the literature (see Table \ref{smg_table}). We restrict this comparison to SMGs at z $< 3$ so that the observed 850$\mu$m flux measures will be longward of the far infrared peak 
in the rest frame. 

We also restrict our sample to only those SMGs with high signal-to-noise ratio (SNR) measurements of CO(1-0). From these, we can reliably obtain ISM mass estimates without large excitation 
corrections and uncertainties. Although many galaxies have detectable emission in higher CO transitions, this high excitation gas, very likely constitutes 
just a small fraction of the total molecular gas. The bulk of the molecular gas is probably at 10 - 20 K and therefore emits appreciably only in the first couple CO lines. And, the CO(1-0) line luminosity 
is well known to be linearly correlated with the gravitational masses of GMC \citep{sco87,sol87}. Although some of these SMGs are likely to be lensed, we do not 
expect a differential magnification difference between the cold molecular gas and the cold dust emission; hence, their ratios should still provide a check on the high redshift luminosity-to-mass 
ratio. 

\begin{figure}[ht]
\epsscale{1.}  
\plotone{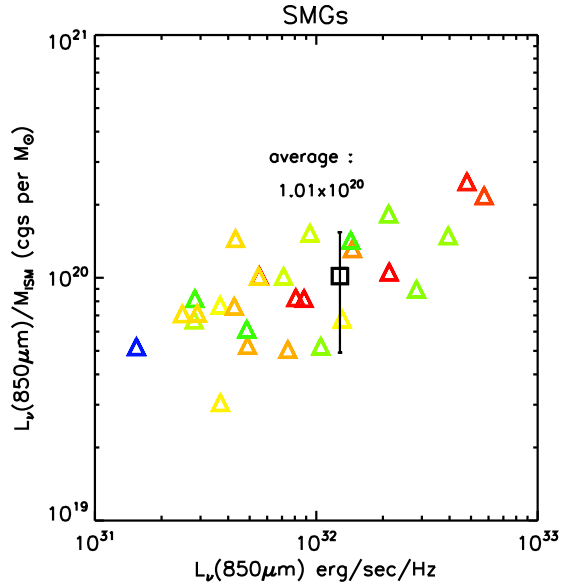}
\caption{The ratio of $ L_{\nu}$ at 850\,$\mu$m to $M_{\rm ISM}$ is shown for a
sample of $z < 3$ SMGs with CO(1-0) data (references given in Table \ref{smg_table}). Colors of the symbols indicate the redshifts ranging 
from z = 1.44 (blue) to 3.0 (red). The average value of the luminosity-to-mass ratio = $1.01\pm0.52 \times 10^{20}$ erg cm$^{-2}$ sec$^{-1}$ \msun$^{-1}$ is indicated by the square. 
(The median value is $8 \times 10^{19}$ erg cm$^{-2}$ sec$^{-1}$ \msun$^{-1}$.) }
\label{smg} 
\end{figure}

The molecular ISM masses were derived from the observed CO(1-0) line fluxes using Eq. 3 in \cite{sol05} 
and a standard $\alpha_{CO} = M_{ISM} / L_{CO} = 4.6$
\msun (K km sec$^{-1} pc^2)^{-1}$. Columns 9 and 10 list the derived ISM masses and $L_{\nu_{850\mu \rm m}} $. The individual galaxy ratios are given in the last column and shown 
in Figure \ref{smg} 
along with the sample average
\begin{eqnarray}
  <\alpha_{850\mu \rm m}>_{SMG} &=& 1.0\pm0.5\times10^{20}  \nonumber  \\
  &&   ~~~~~\rm ergs/sec/Hz/\msun. \label{alpha}
\end{eqnarray}

\begin{deluxetable*}{lrcrrcccccccc}[ht]\label{smg_table}
\tablecaption{SMGs with CO(1-0)  data}
\tablewidth{0pt}
\tablehead{  \colhead{SMG} & \colhead{Ref.} & \colhead{z} &  \colhead{S$_{\nu}(850\,\mu)$} &\colhead{SNR$_{850\mu m}$\tablenotemark{a}}&  \colhead{${\Gamma_{0} \over{\Gamma_{RJ} }}$} & \colhead{$I_{CO}$} & \colhead{SNR$_{CO}$\tablenotemark{a}} & \colhead{$M_{\rm ISM}$\tablenotemark{b}} &  \colhead{$L_{\rm 850}$} & \colhead{$L_{\rm 850}$ / M$_{ISM}$\tablenotemark{b}}\\
\colhead{}  & \colhead{}  & \colhead{} & \colhead{mJy} & \colhead{ } & \colhead{ } & \colhead{Jy km s$^{-1}$}  &  \colhead{}   & \colhead{$10^{11}$ \msun} & \colhead{$10^{31}$cgs} & \colhead{$10^{20}$ cgs/\msun}}
\startdata
\\
      HXMM01... &   1 & 2.31 &   27.0 &   9.0 &    2.6 &  1.73 &    5.6 &   20.1$\pm$   3.6 &   10.5 &    0.5$\pm$   0.2 \\
SPT-S053816-... &   2 & 2.79 &  125.0 &  17.9 &    3.2 &  1.20 &    6.0 &   19.3$\pm$   3.2 &   47.9 &    2.5$\pm$   0.6 \\
HATLASJ08493... &   3 & 2.41 &   19.0 &   9.5 &    2.7 &  0.56 &    8.0 &    7.0$\pm$   0.9 &    7.1 &    1.0$\pm$   0.2 \\
H-ATLASJ0903... &   4 & 2.31 &   54.7 &  17.6 &    2.6 &  1.00 &    7.7 &   11.6$\pm$   1.5 &   21.2 &    1.8$\pm$   0.3 \\
H-ATLASJ0913... &   4 & 2.63 &   36.7 &   9.4 &    3.0 &  0.76 &    6.3 &   11.1$\pm$   1.7 &   14.6 &    1.3$\pm$   0.3 \\
H-ATLASJ0918... &   4 & 2.58 &   18.8 &  11.8 &    2.9 &  1.04 &    4.0 &   14.7$\pm$   3.7 &    7.4 &    0.5$\pm$   0.2 \\
     HLSW-01... &   5 & 2.96 &   52.8 & 105.6 &    3.5 &  1.14 &   10.4 &   20.2$\pm$   2.0 &   21.4 &    1.1$\pm$   0.1 \\
H-ATLASJ1132... &   4 & 2.58 &  106.0 &   5.9 &    2.9 &  0.66 &    3.5 &    9.3$\pm$   2.7 &    4.9 &    0.5$\pm$   0.2 \\
H-ATLASJ1158... &   4 & 2.19 &  107.0 &   5.9 &    2.5 &  0.74 &    6.2 &    7.9$\pm$   1.3 &    4.9 &    0.6$\pm$   0.2 \\
H-ATLASJ1336... &   4 & 2.20 &   36.8 &  12.7 &    2.5 &  0.93 &    7.8 &   10.0$\pm$   1.3 &   14.3 &    1.4$\pm$   0.3 \\
H-ATLASJ1344... &   4 & 2.30 &   73.1 &  30.5 &    2.6 &  2.74 &    7.0 &   31.7$\pm$   4.5 &   28.4 &    0.9$\pm$   0.2 \\
H-ATLASJ1413... &   4 & 2.48 &   33.3 &  12.8 &    2.8 &  1.47 &    8.6 &   19.4$\pm$   2.2 &   13.1 &    0.7$\pm$   0.1 \\
SMMJ2135-010... &   6 & 2.33 &  106.0 &   8.8 &    2.6 &  2.25 &    9.8 &   26.5$\pm$   2.7 &   39.4 &    1.5$\pm$   0.3 \\
SPT-S233227-... &   2 & 2.73 &  150.0 &  13.6 &    3.1 &  1.70 &    6.8 &   26.4$\pm$   3.9 &   57.3 &    2.2$\pm$   0.5 \\
SMMJ123549.4... &   7 & 2.20 &    8.3 &   3.3 &    2.5 &  0.32 &    8.0 &    3.4$\pm$   0.4 &    2.8 &    0.8$\pm$   0.3 \\
SMMJ123707.2... &   7 & 2.49 &   10.7 &   4.0 &    2.8 &  0.91 &    7.0 &   12.1$\pm$   1.7 &    3.7 &    0.3$\pm$   0.1 \\
SMMJ163650.4... &   7 & 2.38 &    8.2 &   4.8 &    2.7 &  0.34 &    8.5 &    4.2$\pm$   0.5 &    2.8 &    0.7$\pm$   0.2 \\
SMMJ163658.1... &   7 & 2.45 &   10.7 &   5.3 &    2.8 &  0.37 &    5.3 &    4.8$\pm$   0.9 &    3.7 &    0.8$\pm$   0.3 \\
EROJ164502+4... &   9 & 1.44 &    4.9 &   6.6 &    1.8 &  0.60 &    6.0 &    3.0$\pm$   0.5 &    1.5 &    0.5$\pm$   0.2 \\
SMMJ02399-01... &  10 & 2.81 &   23.0 &  12.1 &    3.3 &  0.60 &    5.0 &    9.8$\pm$   2.0 &    8.1 &    0.8$\pm$   0.2 \\
SMMJ04135+10... &  10 & 2.85 &   25.0 &   8.9 &    3.3 &  0.64 &    7.9 &   10.7$\pm$   1.4 &    8.8 &    0.8$\pm$   0.2 \\
SMMJ04431+02... &  11 & 2.51 &    7.2 &   4.8 &    2.8 &  0.26 &    4.3 &    3.5$\pm$   0.8 &    2.5 &    0.7$\pm$   0.3 \\
SMMJ14009+02... &  10 & 2.93 &   15.6 &   8.2 &    3.5 &  0.31 &   15.5 &    5.4$\pm$   0.3 &    5.5 &    1.0$\pm$   0.2 \\
SMMJ14011+02... &  10 & 2.57 &   12.3 &   7.2 &    2.9 &  0.40 &    8.0 &    5.6$\pm$   0.7 &    4.3 &    0.8$\pm$   0.2 \\
SMMJ163555.2... &  10 & 2.52 &   12.5 &  15.6 &    2.9 &  0.22 &    5.5 &    3.0$\pm$   0.5 &    4.3 &    1.4$\pm$   0.4 \\
SMMJ163554.2... &  12 & 2.52 &   15.9 &  22.7 &    2.9 &  0.40 &   10.0 &    5.4$\pm$   0.5 &    5.5 &    1.0$\pm$   0.1 \\
SMMJ163550.9... &  12 & 2.52 &    8.4 &  10.5 &    2.9 &  0.30 &    3.3 &    4.1$\pm$   1.2 &    2.9 &    0.7$\pm$   0.3 \\
HATLASJ08493... &  12 & 2.41 &   25.0 &  12.5 &    2.7 &  0.49 &    8.2 &    6.1$\pm$   0.8 &    9.4 &    1.5$\pm$   0.3 \\
\\
 & & &  & & & & & {\bf average\tablenotemark{c}}&{\bf :} & {\bf 1.01 $\pm$ 0.52} \\
\\
\enddata
\tablecomments{Submm fluxes and CO(1-0) measurements from references given in the second column:
1:\citep{fu13}, 2:\citep{ara13}, 3:\citep{ivi13}, 4:\citep{har12}, 5:\citep{rie11}, 6:\citep{les11}, 7:\citep{ivi11}, 8:\citep{rie11a}, 9:\citep{gre03}, 10:\citep{tho12}, 11:\citep{har10}, 12:\citep{ivi13},\citep{bus13}}
\tablenotetext{a}{SNR (signal to noise ratio) calculated from the ratio of the measured flux to stated uncertainty in the observational reference.}
\tablenotetext{b}{Specified uncertainties in the ISM masses and the luminosity-to-mass ratios include only the flux uncertainties for the CO and 850$\mu$m measurements. The uncertainty due to variations in the CO to H$_2$ conversion ratio are difficult to quantify since the physical conditions in the galaxies are unresolved. It is worth noting that for a large sample of Galactic GMCs with both CO (1-0) luminosities and virial masses, \cite{sco87} measured a dispersion of only 7\% in the ratio $L_{CO}/M_{vir}$ and this sample including clouds with and without HII regions and covered a mass range $1-16\times10^5$\msun.}
\tablenotetext{c}{Unweighted average and standard deviation.}
\label{smg_table}
\end{deluxetable*}

In the above analysis we used the standard Galactic factor $\alpha_{CO}$ to convert observed CO(1-0) luminosity to gas mass. As discussed in 
\cite{sol05}, low-z studies of ULIRGs have led to the suggestion that the conversion factor could be several times smaller \citep{dow93,bry99}. In the ULIRGs 
the gas is concentrated in the nuclear regions as a result of dissipative galaxy merging and the molecular gas line width can be increased 
by the galactic dynamics, including the stellar mass -- not just the gas mass as in the self-gravitating GMCs where the standard conversion 
factor was derived. In addition, the mean gas temperature and density ($\rho$) may be different in the ULIRG nuclei as a result of the intense 
star formation activity and the $\alpha_{CO}$ should vary as $<\rho>^{1/2}/T_k$ \citep[see Equation 8.5 in][]{sco12}. 

For the high z SMGs, it is not obvious that 
the lower $\alpha_{CO}$ (used in low-z ULIRGs) is appropriate since it is uncertain that the bulk of the molecular gas in the SMGs is similarly concentrated. 
Our restriction to  CO(1-0) in the above sample was specifically intended to avoid sensitivity to the presence of high excitation 
gas, and to sample the larger, presumably extended masses of cold gas. Indeed, the ratio of dust emission to 
gas mass is similar to that obtained in low z galaxies; this suggests that the SMG CO(1-0) emission is not enhanced by 
concentration in the galactic nuclei.

\subsection{Summary -- an approximately constant RJ mass-to-light ratio}

In the preceding sections, we have presented the physical explanation and, more importantly, strong 
empirical justification for using the long wavelength RJ dust emission in galaxies as a linear 
probe of the ISM mass. The most substantial determination of the dust RJ spectral slope is that 
obtained by the Planck Mission from observations of the Galaxy \citep{pla11a,pla11b}, indicating a dust emissivity index 
$\beta = 1.8 \pm 0.1$ with no strong evidence of variation in Galactic radius or between atomic and molecular regions. 
Secondly, both the Planck data and measurements for nearby local galaxies, including both normal star forming and star bursting systems, 
indicate a similar constant of proportionality $\alpha_{850}$ for the dust emission at rest frame 850$\mu$m per unit mass of ISM. Lastly, 
we find for a large sample of SMGs at z = 1.4 to 3, their ratio of rest frame 850$\mu$m per unit mass of ISM is essentially identical to that obtained 
for the local galaxies. The similarities of the $\alpha_{850}$s argue strongly that for the majority of ISM mass, the dust emissivity at long wavelengths, 
the dust-to-gas mass ratio and the dust temperatures vary little. 
The dispersions (or uncertainties) in the derived $\alpha_{850\mu m}$ and $\beta$ are
5 - 25\%. Since these uncertainties are generally less than the flux measurement fractional uncertainties we include only the latter in the following 
uncertainty estimates.

The submm flux to dust mass ratio is expected to vary linearly with dust temperature. In practice, the overall range of $T_d$ for the bulk of the 
mass of ISM is very small, since it requires very large increases in the radiative heating to increase the dust temperatures ($T_d$ varies approximately 
as the 1/5 - 1/6 power of the radiation energy density). As noted above, the extensive surveys of local galaxies with Herschel find a 
range of $T_d \sim 15 - 30$ K \citep{dun11,dal12,aul13}. Where we have needed to specify a dust temperature (e.g. for the R-J correction)
we used 25 K, so we expect the uncertainties in the derived masses averaged on galaxy scales will be less than $\sim 25$\%. 

These calibrations, including atomic- and molecular-dominant regions; normal to star bursting systems; inner to outer galaxy; 
and low to moderate redshift lay a solid foundation for using measurements of the RJ dust emission to probe galactic ISM masses. 
ALMA enables this technique for high redshift surveys, providing high sensitivity and the requisite angular resolution to avoid source confusion.

\subsection{Two Cautions}

It is important to recognize that even for those objects detected in SPIRE, the SPIRE data can not be used to reliably estimate ISM masses (along the 
lines as done here) for the z = 1 and 2 samples. For those redshifts, the SPIRE data will be probing near the rest frame far infrared luminosity peak -- {\it not safely on the RJ tail and not necessarily 
optically thin}.
The longest wavelength channel (500$\mu$m or 600 GHz) will be probing rest frame 170$\mu$m for z = 2; for such measurements, there will be substantial uncertainty in the mass estimate, depending on the assumed value of the dust temperature (see Figure \ref{rj}). 

\emph{Fitting the observed SED to derive an effective dust temperature is {\bf not} a reliable approach} -- near the far infrared peak, the temperature characterizing the emission is 'luminosity-weighted' (i.e. grains undergoing strong radiative heating) rather than mass-weighted. Hence, the derived $T_d$ will not reflect the temperature appropriate to the bulk of the ISM mass. Or put another way, the flux measured near the peak is simply a measure of luminosity -- not mass. Furthermore, the 
large SPIRE beam results in severe source confusion at these flux levels and hence unreliable flux measurements for individual galaxies. At z $> 2$ ALMA resolution and sensitivity 
are required and one must observe at $\nu \leq 350$ GHz to be on the RJ tail of the dust emission.

\section{COSMOS Mass-Selected Galaxy Sample for ALMA}\label{sample}

\begin{deluxetable}{ccccc}
\tablecaption{Galaxy Samples for Dust Continuum Measurements  }
\tablewidth{0pt}
\tablehead{  \colhead{Sample } & \colhead{\# } & \colhead{z } & \colhead{$< $M$_{stellar} >$\tablenotemark{a}} &  \colhead{$log< $SFR$ >$\tablenotemark{a}} \\
\colhead{}  & \colhead{} & \colhead{} & \colhead{$10^{11}$\msun }  &  \colhead{\msun yr$^{-1}$} }
\startdata
Low-z &    37   &   0.22 - 0.48    & 1.30$\pm$0.57 &   0.63$\pm$1.72   \\
Mid-z &    33   &   0.81 - 1.15    & 1.59$\pm$0.66 &   1.31$\pm$1.68           \\
High-z &    31   &   1.46 - 2.54    & 1.23$\pm$0.61 &   1.97$\pm$0.57    \\
IR-Bright &  6   &   1.46 - 2.02    & 1.00$\pm$0.33 &    2.27$\pm$0.23      \\
\enddata \label{samples}
\tablenotetext{a}{The uncertainties give the dispersion of each sample.}
\end{deluxetable}

Our sample of 107 galaxies is taken from the COSMOS 2 deg$^2$ survey \citep{sco_ove} which has  excellent photometric redshifts derived from 34 band (UV-Mid IR) photometry. The COSMOS field also has deep infrared coverage with Spitzer (IRAC and MIPS) and Herschel (PACS and SPIRE),  as well as radio continuum and X-ray coverage \citep{ilb13}. 

\begin{figure}[ht]
\epsscale{1.}  
\plotone{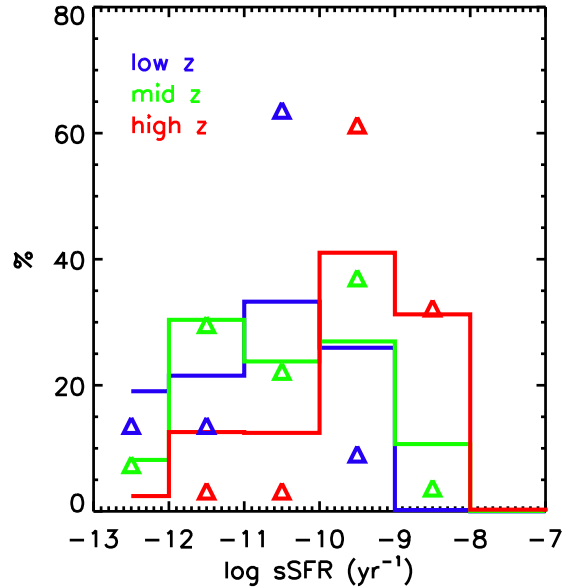}
\caption{ The distribution of specific star formation rates are shown for the three redshift samples. The histograms show the distributions 
obtained from the full COSMOS photometric redshift catalog using the same redshift ranges and stellar mass ranges as observed here. The triangle 
points are the distributions for the three galaxy samples observed here. 
}
\label{sample_ssfr} 
\end{figure}

The galaxies were selected to have stellar masses 
$M_* \simeq10^{11}$ \msun, as determined from the photometric redshift SED template fit.  The photometric redshifts and stellar masses of the galaxies are from \cite{ilb13} and the accuracy of those quantities is discussed in detail there. The SFRs are derived from the rest frame UV continuum and the infrared using Herschel PACS and SPIRE data as detailed in \cite{sco13}. For half of the objects there exist reliable (multi-line) spectroscopic redshifts.

In order to span the range of star formation activity for massive galaxies at each redshift, the galaxies were selected to span the range of NUV - r colors at each redshift.  In the lower redshift samples, an increasingly large fraction of the objects would be classified as "passive" based on their
red NUV - r colors. Figure \ref{sample_ssfr} shows the distributions of specific star formation rates ($sSFR = SFR / M_*$) for the full COSMOS galaxy 
samples and for the samples observed here. 

The properties of the individual galaxies and their ALMA flux measurements and upper limits are listed in Tables \ref{lowz} - \ref{highz} in \S \ref{app}. 
The observed galaxies fall into four subsamples, all stellar mass-selected but over a range of redshifts. Their summary properties are given in Table \ref{samples} including the 
median stellar mass (M$_*$) and star formation rate (SFR) of each sample. The first three samples were chosen based purely on stellar mass with no prior selection 
for either SFR or far infrared luminosity.  The last sample was specifically chosen to be infrared bright (easily detected by Herschel) and to have stellar masses $\sim10^{11}$ \msun, like the other samples.

\section{Observations and Results}\label{obs}

The ALMA Cycle 0 observations for project (\#2011.0.00097.S) analyzed here were obtained in Mar - Oct 2012 in Band 7 at 350 GHz. On-source integration times were 1, 2 and 4 minutes per galaxy 
at z  = 0.3, 0.9 and 2, respectively. With continuum bandwidths of 8 GHz, the 1$\sigma$ rms sensitivity was 0.5, 0.4 and 0.3 mJy and typical synthesized beam sizes 
were $\simeq 0.6$\arcsec. 
The data were calibrated and imaged with natural weighting using CASA.

\subsection{Source Measurement and Noise Estimation}

The measurement results for the individual galaxies are tabulated in Tables \ref{lowz} -- \ref{highz}.  In the images, apertures centered on the
galaxy position were used to search for significant detection of continuum in the integrated aperture flux (S$_{tot}$ in the tables) or a 
high single pixel (S$_{peak}$ in the tables) within the aperture. The former recovers instances where the emission is significantly extended beyond the synthesized beam ($\simeq 1$\arcsec); the latter is most sensitive when the emission is unresolved. The apertures were 3 and 2 \arcsec for the low z sample and the mid and high z samples respectively. A significant detection required a 2$\sigma$ SNR in the integrated aperture flux (S$_{tot}$ in the tables) or a 3.6 $\sigma$ detection of a 
high single pixel (S$_{peak}$ in the tables) within the aperture. These different n$\sigma$ limits 
were chosen such that there would be less than one spurious detection by either technique in the entire sample.

The noise estimate in both cases was derived from the dispersion in the integrated and peak flux measurements obtained 
for 100 apertures of the same size displaced off-center in the same image. We also tabulate the signal-to-noise ratios
(SNRs) given by

  \begin{eqnarray} \label{snr}
  \rm{SNR_{tot}}  &=& \rm{{S_{tot} \over{ \sigma_{tot} } }} \nonumber \\
   \rm{SNR_{peak} } &=& \rm{{S_{peak} \over{ \sigma_{pix} } }}
 \end{eqnarray}

\noindent Note that we let the SNR be less than 0 if the measured flux is less than 0; this is so that large negative flux values don't end up with a positive SNR (above the detection thresholds).

The signal-to-noise ratio given in Column 9 (Tables \ref{lowz} --- \ref{highz}) is the better of those obtained
from the integrated or peak flux measurement. In the last column, limits are given at 2$\sigma$ and 3.6$\sigma$ in the inferred mass,
depending on whether the better SNR was obtained for the integrated or peak flux measurement.

Twenty-eight of the 107 galaxies (26\%) were significantly detected and these are shown in Figure \ref{detected}. The maximum detected 350 GHz 
flux was $\sim 4$mJy and the inferred ISM masses are $1 - 8 \times 10^{10}$ \msun. Although the detection rate is less than 50\%, one should recall that the integration 
times were only 1 - 4 minutes with fewer than 18 of the eventual 64 ALMA telescopes. Secondly, this very high stellar mass sample will preferentially include 
a large fraction of passive galaxies and the ISM contents may be larger at lower stellar mass. The detection rate was 100\% for the IR bright sample. 

 \section{Stacked Samples}
 
 \begin{figure}[ht]
\epsscale{1.}  
\plotone{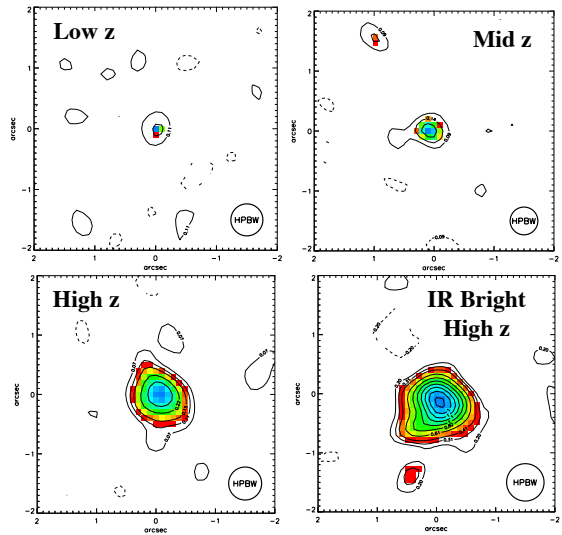}
\caption{ Median stacked images for the four subsamples of galaxies (Tables \ref{samples} and \ref{stacks}).
The scale is in arcsecs and the contours are at -2, 2, 3, 4, 5, 7, 9, 11 $\sigma$ (see Table \ref{stacks}).}
\label{stack_images} 
\end{figure}

\begin{figure}[ht]
\epsscale{1.}  
\plotone{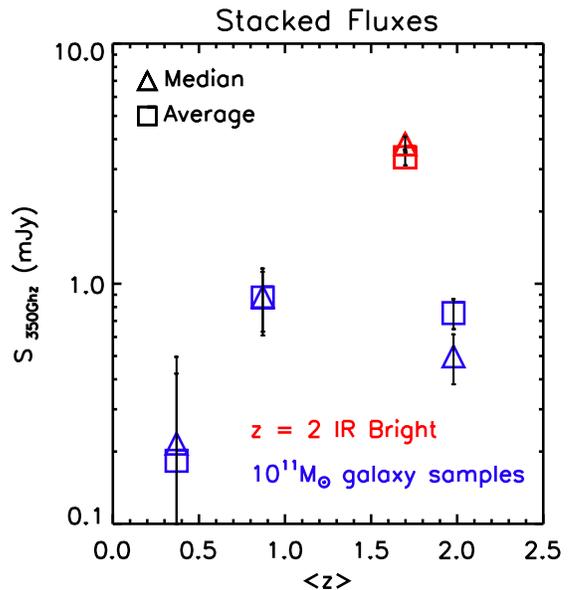}
\caption{ The 350 GHz continuum fluxes from stacked images of the four galaxies samples. Median and average stacked images are indicated
by triangles and squares. The signal-to-noise ratios, SNRs, are given in Table \ref{stacks} and they are from $\sim$3, 4, 8 and 16 for the 
low-z, mid-z, high-z and IR bright samples, respectively. 
}
\label{flux_stack} 
\end{figure}

\begin{figure*}[ht]
\epsscale{1.}  
\plottwo{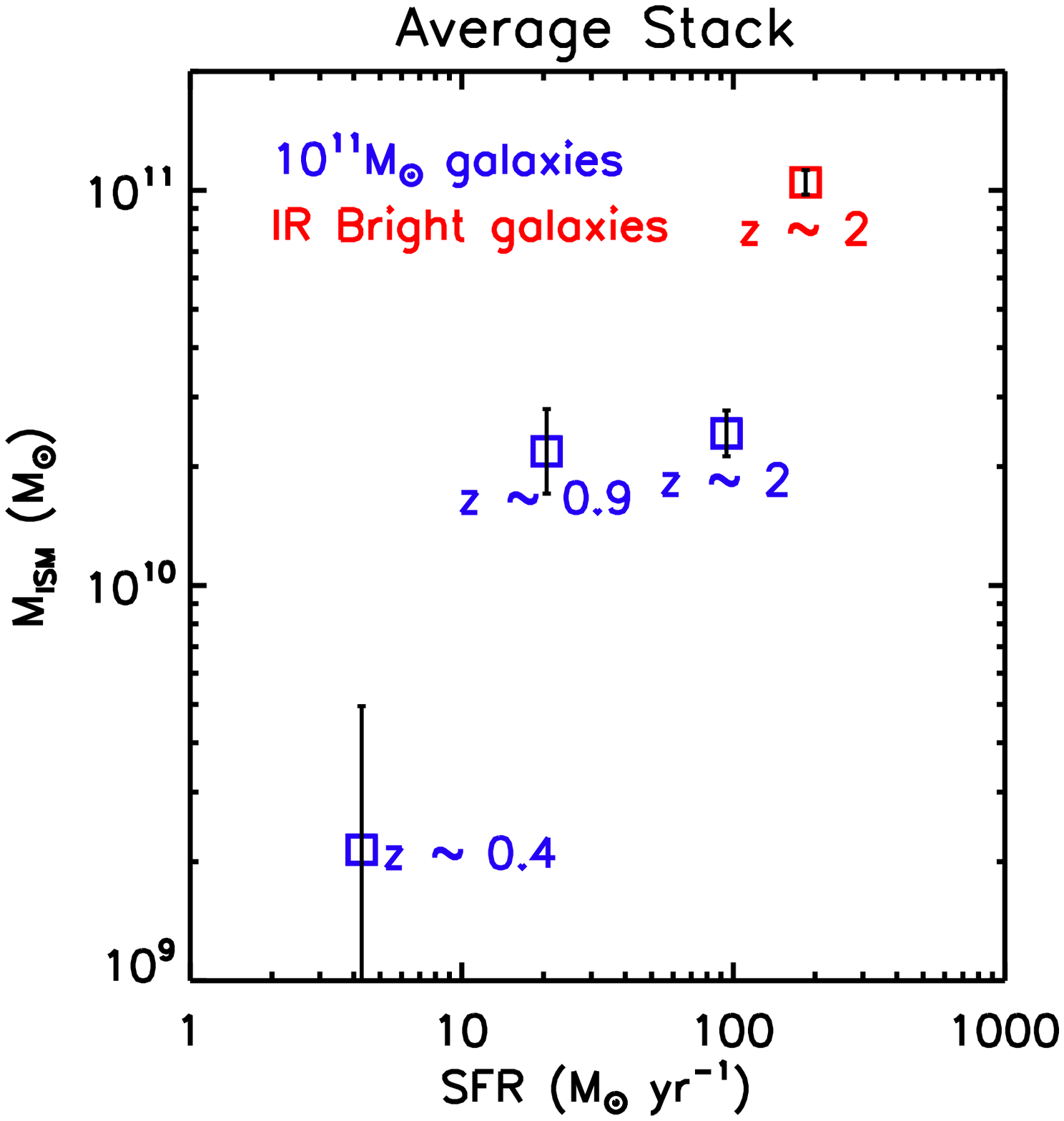}{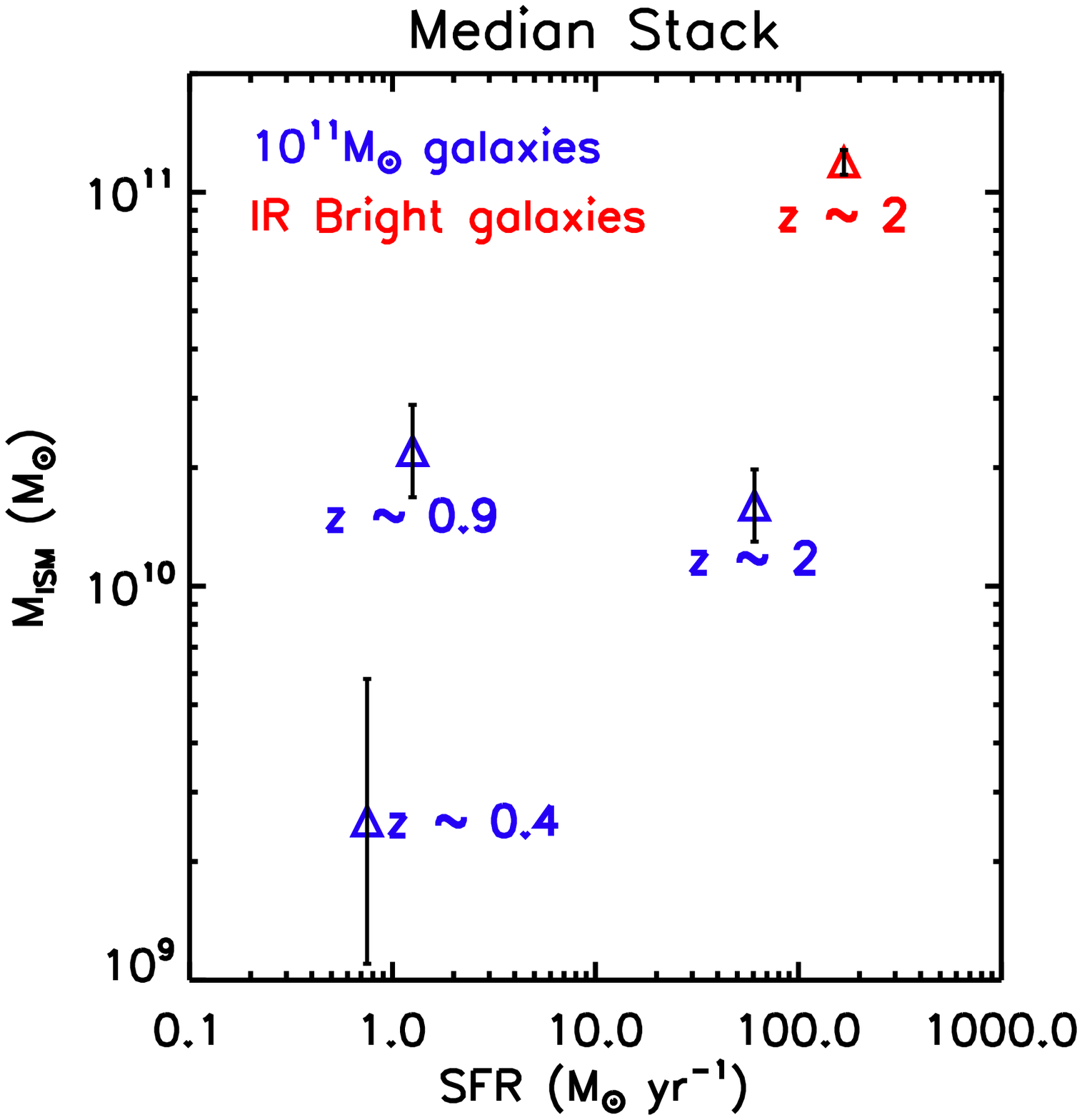}
%\plotone{dust_measures_plot_p6.pdf}
%\plotone{dust_measures_plot_p8.pdf}
\caption{Derived ISM masses of the average (left) and median (right) stack images are shown with their median and average SFRs. 
The star formation timescales ($\tau_{SF} = M_{ISM} / <SFR>$), listed in Table \ref{stacks} are in the range 0.2 to 1.2 Gyr.}
\label{ism_stack} 
\end{figure*}

The mean fluxes and derived ISM masses for stacked images of each of the four galaxy subsamples may be used to explore the 
overall evolution of the dust emission and ISM masses. The individual galaxy images were stacked 
in three ways: an average, a median stack and an average weighted by the square of the rms noise. The median and weighted averages are used in the discussion and figures
below. [The simple average was not significantly different except that it has higher noise than the rms-weighted average.]  Figure \ref{stack_images} shows the 
median stacked images for each of these subsamples.

\begin{deluxetable*}{ccccccccccc}[ht]
%\tabletypesize{0.2}
\tablecaption{\bf{Stacked Samples}  }
\tablewidth{0pt}
\tablehead{ 
 \colhead{stack} &  \colhead{S${\nu}$}  & \colhead{$\sigma_{pix}$} & \colhead{$\sigma_{total}$} & \colhead{SNR\tablenotemark{a}}  & \colhead{$< z >$} &  \colhead{$<$ M$_{*}$$ >$} &  
 \colhead{$log< $SFR$ >$} &   \colhead{$< $M$_{\rm ISM}$$ >$} &   \colhead{$< \tau_{SF}>$} &   \colhead{$M_{ISM}/(M_{*}+M_{ISM})$}\\ 
 \colhead{} & \colhead{mJy} & \colhead{mJy} & \colhead{mJy}  & \colhead{} & \colhead{}  &  \colhead{ $10^{11}$\msun} &  \colhead{\msun yr$^{-1}$}& \colhead{$10^{10}$\msun} & \colhead{Gyr} & \colhead{}}
\startdata 
\\
& & & & \bf{Low-z} \\
   weighted average   &      0.18 &     0.057 &     0.240 &      3.23 &      0.39 &  1.30$\pm$0.57 &      0.63$\pm$1.72 &  0.21$\pm$0.07 &  0.50$\pm$0.16 &  0.02$\pm$0.005 \\
       median         &      0.22 &     0.073 &     0.280 &      3.55 &      0.37 &  1.11$\pm$0.57 &     -0.13$\pm$1.72 &  0.25$\pm$0.07 &  0.64$\pm$0.18 &  0.02$\pm$0.006 \\
       \\
& & & & \bf{Mid-z} \\
   weighted average   &      0.88 &     0.045 &     0.245 &      4.63 &      0.89 &  1.59$\pm$0.66 &      1.31$\pm$1.68 &  2.19$\pm$0.47 &  1.06$\pm$0.23 &  0.12$\pm$0.026 \\
       median         &      0.88 &     0.059 &     0.274 &      3.51 &      0.87 &  1.41$\pm$0.66 &      0.10$\pm$1.68 &  2.20$\pm$0.63 &  1.15$\pm$0.33 &  0.13$\pm$0.038 \\
       \\
& & & & \bf{High-z} \\
   weighted average   &      0.75 &     0.037 &     0.108 &      9.96 &      2.07 &  1.23$\pm$0.61 &      1.97$\pm$0.57 &  2.43$\pm$0.24 &  0.26$\pm$0.03 &  0.16$\pm$0.017 \\
       median         &      0.50 &     0.047 &     0.117 &      6.72 &      1.98 &  1.09$\pm$0.61 &      1.78$\pm$0.57 &  1.60$\pm$0.24 &  0.16$\pm$0.02 &  0.13$\pm$0.019 \\
       \\
& & & & \bf{IR bright} \\
   weighted average   &      3.36 &     0.102 &     0.249 &     18.44 &      1.66 &  1.00$\pm$0.33 &      2.27$\pm$0.23 & 10.48$\pm$0.57 &  0.57$\pm$0.03 &  0.51$\pm$0.028 \\
       median         &      3.82 &     0.130 &     0.286 &     15.42 &      1.70 &  0.93$\pm$0.33 &      2.22$\pm$0.23 & 11.91$\pm$0.77 &  0.54$\pm$0.04 &  0.56$\pm$0.036 \\
  \\ 
\enddata\label{stacks}
\tablecomments{For the mid-z and high-z stacks, the total aperture flux was used. For the low-z stack the flux is from the peak pixel flux within the 3\arcsec aperture. This was necessitated since the lower 
signal strength did not allow a significant detection in the integrated aperture flux. $\tau_{SF} = M_{ISM} / <SFR>$ where $<SFR>$ is the mean SFR. The uncertainties given for each quantity are the standard deviations.}
\tablenotetext{a}{SNR$_{tot}$ and SNR$_{peak}$ calculated separately from Eq. \ref{snr}  and the column SNR lists the larger in absolute magnitude of those two SNRs. }
\end{deluxetable*}

Figure \ref{flux_stack} shows the average and median 350 GHz fluxes for each of the four stacked samples. Table \ref{stacks} lists these measurements along 
with their signal-to-noise ratios. All four samples are significantly detected in the stacked images (both median and average). The lowest redshift 
sample does not have high statistical significance in the stacked detection but that is to be expected, given the anticipated evolution 
of ISM masses for these very massive galaxies. 

The derived ISM masses for each of the samples are shown in Figure \ref{ism_stack} for the average and median stacks.  The figures clearly 
show an increase in the ISM content both in absolute mass and gas mass fraction (since the stellar masses are the same) from z = 0.4 
to 1. From z = 1 to 2, the ISM mass shows no significant evolution for the strictly stellar mass-selected samples. This is a 
very interesting result, but given the small sample sizes and the fact that these are only the most massive galaxies, it is not clear 
how general these results are in terms of overall cosmic evolution. A high priority should be placed on extending the 
samples to lower stellar mass and better sampling of the star formation characteristics. 

Comparing the IR bright sample at z  =  2 with the strictly mass-selected sample at the same redshift, it is clear that the IR bright galaxies 
have considerably higher long wavelength fluxes and presumably higher ISM contents by a factor $\sim 5$.  This increase in the apparent ISM mass 
is in fact larger than the increase in the estimated SFRs. If this holds up, it would clearly suggest that the z = 2 IR bright galaxies 
are extra luminous simply because they have more ISM. 

The depletion timescales for the ISM via star formation are given by $\tau_{SF} = M_{ISM} / <SFR>$ and are also listed in Table \ref{stacks}. This depletion timescale is similar for the subsamples (0.2 - 1.2 Gyr). Both of the z $\sim 2$ samples have similar depletion timescales (0.2 and 0.5 Gyr for the normal and IR bright samples, respectively). This underscores the point made above that the IR bright galaxies simply have more
ISM, rather than having a higher star formation efficiency. 

The gas mass fractions ($M_{ISM}/(M_{*}+M_{ISM})$) are $\sim 2\pm0.5, 12\pm3, 14\pm3 ~\rm{and} ~53\pm5$ \% for the 
low-z, mid-z, high-z and IR bright samples, respectively.  \cite{tac13} surveyed 52 galaxies at z = 1 - 3 in CO(3-2) and found mean gas mass fractions 
of 0.33 and 0.47 at z $\sim 1.2$ and 2.2, respectively.  Their sample was selected with $M_* > 2.5 \times 10^{10}$ \msun and SFR $> 32$ \msun yr$^{-1}$; it is therefore 
likely to favor more active star forming galaxies than our purely stellar-mass selected samples. On the other hand, our IR-bright sample is likely to 
be more star forming than their average galaxy so it is entirely reasonable to find a somewhat higher gas mass fraction.

\section{Summary}

We have developed the physical and empirical basis for using the long wavelength RJ dust emission as an accurate 
and fast probe of the ISM content of galaxies.

To obviate the need to know both the dust opacity and dust-to-gas ratio (which are degenerate in Eq. \ref{fnu}), we have 
chosen instead to empirically calibrate  
the ratio of the specific luminosity at rest frame 850$\mu$m to total ISM mass using samples of observed galaxies, thus absorbing the opacity curve and the abundance ratio into a single empirical constant $ \alpha_{850\mu \rm m} =  L_{\nu_{850\mu \rm m}} / M_{\rm ISM} $.
Three samples were developed: 1) 12 local star forming and starbursting galaxies with global SCUBA and ISM measures; 
2) extensive Galactic observations from the Planck Collaboration and 3) a sample of 28 SMGs at z $< 3$ having CO(1-0) measurements. The 3 samples yielded  $\alpha_{850\mu \rm m} $= 1.0, 0.79 and 1.01
$\times10^{20}  \rm ~ergs~ sec^{-1} Hz^{-1} \msun^{-1}$, respectively. 

The Planck measurements are particularly convincing --  they are of high SNR, span a large wavelength range and they probe the diversity of Galactic ISM
including both HI and H$_2$ dominated clouds,  exhibiting little variation in the empirical $\alpha_{250\mu \rm m}$  \citep{pla11a}. The Planck measurements also determine very well the long wavelength dust emissivity index: $\beta = 1.8 \pm 0.1$ \citep{pla11b}. 

In our ALMA observations, we intentionally sample high stellar mass galaxies which should have 
close to solar metallicity \citep{erb06} in order to avoid the issue of possible variations in dust-to-gas ratios at low metallicity. However, we note that \cite{dra07b} see no variation in the 
dust abundance down to $\sim20$\% of solar metallicity and it is reassuring that the low z sample and the high z SMGs  yield similar $\alpha_{850\mu \rm m}$.
Lastly, we note that the consistency of all these calibrations (nearby galaxies, the Milky Way and high z SMGs) is strongly suggestive that the CO-to-H$2$ conversion factor ($\alpha_{CO}$) 
does not vary greatly, when applied to global ISM measurements. 

We then applied the derived calibrations to our ALMA dust continuum observations for a sample of 107 galaxies in COSMOS 
with stellar mass $M_* \simeq 10^{11}$ \msun~ in three redshift ranges, z  = 0.4, 0.9 and 2. Strong evolution is 
seen with the ISM mass fraction decreasing from $53\pm5$\% of the total (ISM+stars) mass at the higher redshifts down 
to $\sim2\pm0.5$\% at the lowest redshift. In addition, a small sample of 6 IR luminous galaxies at z = 2 has $\sim5$ times 
greater ISM mass than the equivalent non-IR bright sample at the same redshift. 

Although these preliminary observations detect only a minority of the individual objects, the mean fluxes and masses obtained
from the stacked data will be most worthwhile in providing {\it a priori} estimates for expected fluxes in future observational 
programs. Thus, required sensitivity and time estimates are more reliable, and sample sizes can be chosen appropriately based on the 
fluxes presented in Figure \ref{alma_obs}. 

This was a "pathfinder" study using ALMA in Cycle 0 with just 5 hrs of time in Band 7 (350 GHz) and a maximum of 18 telescopes. Thus the evolutionary 
trends mentioned above are very preliminary and much larger samples are warranted. It should be emphasized that the technique advocated here is both much faster (by at least a factor 10) than using the standard 
molecular line tracers such as CO; it is also is very likely more robust -- avoiding the uncertainties of variable molecular excitation and the need to 
observe higher rotational transitions at high redshift. 

Lastly, we note that using the dust continuum to probe the ISM content does not
require high accuracy spectroscopic redshifts, and may instead be done on much larger samples with accurate photometric redshifts 
such as those in COSMOS. 

Although, the dust continuum can yield accurate and fast measurements of ISM masses, it is 
important to remember that the spectral line measurements do provide vital information such as dynamical masses, excitation and abundance information -- 
clearly these are not available from a single continuum measurement.

\acknowledgments
%\begin{quote}
We thank the referee for a number of useful suggestions. We also thank Zara Scoville for proof reading the manuscript and Andreas Schruba for helpful comments. 
The Aspen Center for Physics and the NSF Grant \#1066293 is acknowledged for hospitality during the initial writing of this paper.
KSS and KS are supported by the National Radio Astronomy Observatory, which is a facility of the National Science Foundation operated under cooperative 
agreement by Associated Universities, Inc. This paper makes use of the following ALMA data: ADS/JAO.ALMA\# 2011.0.00097.S. ALMA is a partnership of ESO (representing its member states), NSF (USA) and NINS (Japan), together with NRC (Canada) and NSC and ASIAA (Taiwan), in cooperation with the Republic of Chile. The Joint ALMA Observatory is operated by ESO, AUI/NRAO and NAOJ
%\end{quote}

\vfill
\clearpage

\bibliography{scoville_dust}{}

%\clearpage
\appendix 

\section{Individual Galaxies and their Fluxes}\label{app}

In Tables \ref{lowz} - \ref{highz} we list the individual flux measurements, and galaxy properties are summarized for all 107 
galaxies in our survey. The objects are taken from the COSMOS survey field \citep{sco_ove} and the galaxy properties are from the 
latest photometric redshift catalog \citep{ilb13}. This catalog has very high accuracy photometric redshifts based on very deep 34 band photometry, including near infrared photometry from the Ultra-Vista survey. 
See \cite{ilb13} for accuracy of the redshifts and the derived galaxy stellar masses. The SFRs in Column 9 are from \cite{sco13}; they were derived from both the 
UV continuum and the Herschel PACS and SPIRE data. 

Columns 5 and 6 in each table list integrated and peak flux measurements for apertures of 3\arcsec (Low-z and Mid-z) and 2\arcsec (High-z) diameter centered on the galaxy position. The aperture sizes are intended to include most of a galactic disk ($\sim 10$ kpc). The noise estimate in both cases was from the measured dispersion in the integrated and peak flux measurements obtained 
for 100 displaced off-center apertures of the same size in each individual image. The signal-to-noise ratio given in Column 7 is the better of those obtained
from the integrated or peak flux measurement; it is the ratio of the signal in Columns 5 and 6 to the measured noise for Columns 5 and 6. In the last column, limits are given at 2$\sigma$ and 3.6$\sigma$ in the inferred mass,
depending on whether the better SNR was obtained for the integrated or peak flux measurement. The detection thresholds of 2 and 3.6 $\sigma$ are chosen such that the chance of a spurious 
'noise' detection across the entirety of each sample is less than $\sim10$\% (based on the measured noise in each image and for the peak flux measurement, based on the number of pixels).

\begin{figure}[ht]
\epsscale{0.6}  
\plotone{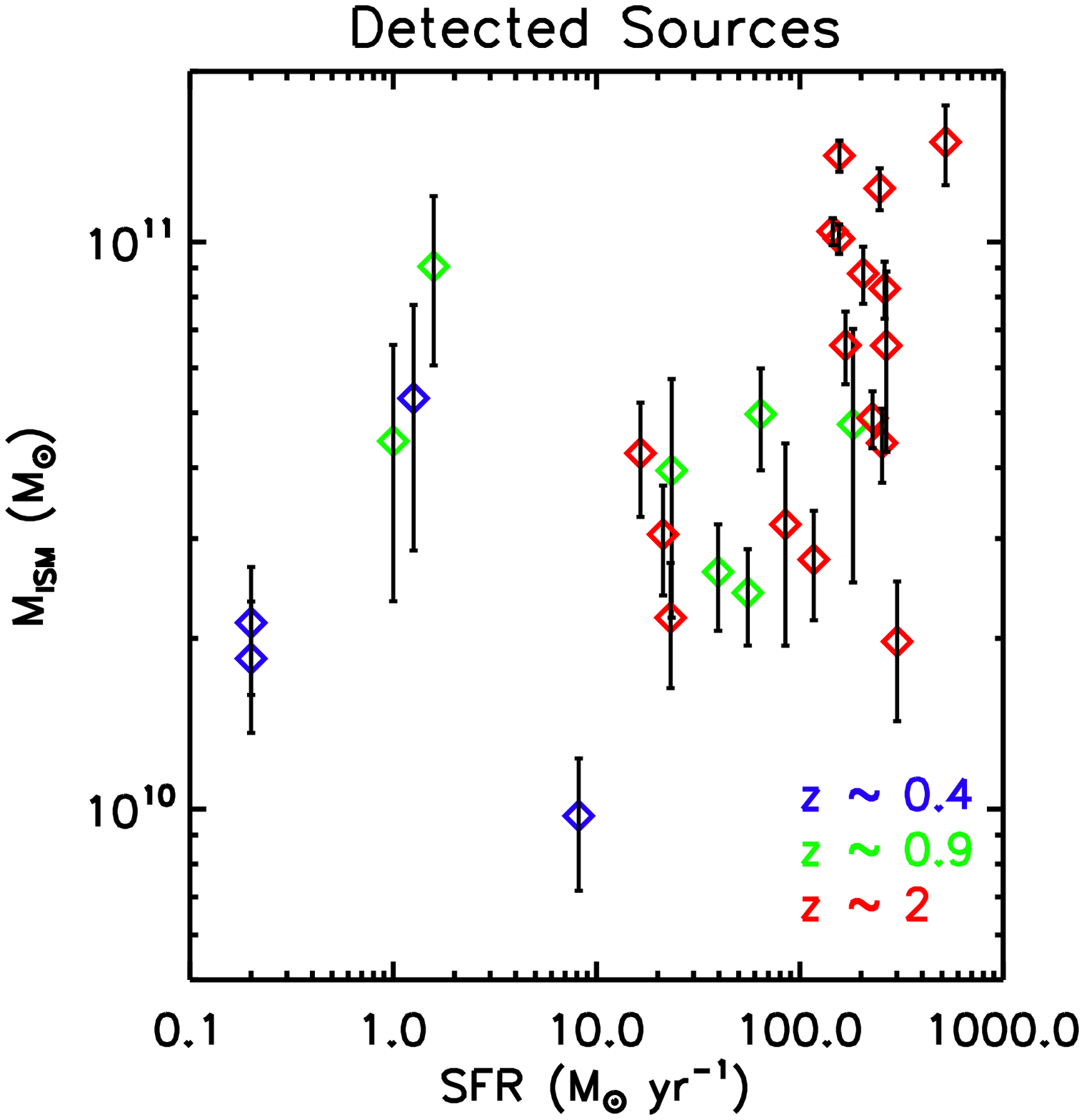}
\caption{Inferred ISM masses are shown for the detected galaxies along with their SFRs derived from the COSMOS rest frame UV continuum combined with
the Herschel PACS and SPIRE fluxes. The redshift bin of each object is indicated by color coding. (For the two lowest SFR sources, their nominal SFRs are actually 10 times
lower than shown but they were plotted at SFR = 0.2 \msun yr$^{-1}$ to maintain a reasonable scale for the other points. Uncertainties in the fluxes and SNRs are give in Tables \ref{lowz} -- \ref{highz}.)
}
\label{detected} 
\end{figure}

For the detected objects, Figure \ref{detected} shows their derived ISM masses based on the dust continuum and their SFRs. 

\begin{deluxetable*}{rrrrrrrrrccr} 
%\tabletypesize{0.2}
%\normalsize
%\rotate
\tablecaption{{\bf Low-z Galaxy Sample in Band 7}} 
\tablewidth{0pt}
\tablehead{ 
 \colhead{\#} &  \colhead{RA (2000)} & \colhead{Dec (2000)} & \colhead{z\tablenotemark{a} } &\colhead{S$_{tot}$} &\colhead{$\sigma_{tot}$} & \colhead{S$_{peak}$} &\colhead{$\sigma_{pix}$} & \colhead{SNR\tablenotemark{b}}  & \colhead{M$_{*}$\tablenotemark{a}} &  \colhead{Log(SFR)\tablenotemark{a}} &   \colhead{M$_{\rm ISM}$}  \\
 \colhead{} & \colhead{\deg} & \colhead{\deg} & \colhead{}  & \colhead{mJy} & \colhead{mJy}  & \colhead{mJy} & \colhead{mJy}  & \colhead{} &  \colhead{ $10^{11}$\msun} &  \colhead{\msun yr$^{-1}$}& \colhead{$10^{10}$\msun} }
\startdata 
& & & & & \bf{Low-z} \\
  1 &   150.2142 &     1.7578 &      0.47 &     -0.06 &      1.57 &      0.91 &      0.44 &     -0.04 &      1.22 &     -3.00 &             $<$ 1.98 \\
  2 &   150.0296 &     1.7375 &      0.28 &     -1.66 &      1.60 &      1.07 &      0.39 &     -1.04 &      0.71 &     -3.00 &             $<$ 0.95 \\
  3 &   149.7943 &     1.7293 &      0.35 &      0.37 &      1.57 &      0.99 &      0.38 &      0.23 &      0.89 &      0.76 &             $<$ 1.21 \\
  4 &   150.4186 &     1.8580 &      0.46 &     -1.92 &      1.18 &      0.73 &      0.27 &     -1.63 &      0.86 &     -1.60 &             $<$ 1.17 \\
  5 &   150.0034 &     1.9173 &      0.27 &      0.85 &      1.04 &      1.05 &      0.42 &      0.82 &      1.40 &     -3.00 &             $<$ 0.96 \\
  6 &   149.8612 &     1.8387 &      0.44 &      1.29 &      1.95 &      1.05 &      0.45 &      0.66 &      1.71 &     -3.00 &             $<$ 1.86 \\
  7 &   149.6918 &     1.8285 &      0.46 &      2.17 &      1.81 &      1.27 &      0.41 &      1.20 &      1.00 &      1.54 &             $<$ 1.79 \\
  8 &   150.5417 &     2.7316 &      0.35 &      1.39 &      0.74 &      0.68 &      0.29 &      1.88 &      3.42 &     -3.00 &             $<$ 0.93 \\
  9 &   150.6447 &     1.9532 &      0.45 &      1.54 &      0.93 &      0.74 &      0.26 &      1.65 &      1.78 &     -3.00 &             $<$ 1.09 \\
 10 &   150.3202 &     2.0114 &      0.31 &     -1.59 &      1.39 &      1.53 &      0.48 &     -1.15 &      0.67 &     -2.71 &             $<$ 1.33 \\
 11 &   150.2115 &     2.0629 &      0.37 &     -0.84 &      2.09 &      1.54 &      0.49 &     -0.40 &      0.79 &     -3.00 &             $<$ 1.68 \\
 12 &   149.6316 &     2.0700 &      0.31 &     -0.81 &      2.19 &      0.98 &      0.43 &     -0.37 &      0.65 &     -2.40 &             $<$ 1.18 \\
 13 &   150.5090 &     2.0401 &      0.37 &      2.07 &      1.05 &      0.83 &      0.28 &      1.98 &      0.90 &     -3.00 &             $<$ 0.97 \\
 14 &   150.0951 &     2.1937 &      0.42 &     -0.08 &      0.89 &      0.65 &      0.27 &     -0.09 &      1.12 &     -0.10 &             $<$ 1.07 \\
 15 &   149.6461 &     1.8434 &      0.37 &     -2.84 &      1.97 &      1.11 &      0.42 &     -1.44 &      0.56 &      0.40 &             $<$ 1.43 \\
 16 &   150.7706 &     2.3259 &      0.45 &     -4.93 &      1.07 &      0.62 &      0.29 &     -4.62 &      1.97 &     -1.14 &             $<$ 1.24 \\
 17 &   150.1714 &     2.2379 &      0.34 &     -1.06 &      2.44 &      1.08 &      0.45 &     -0.43 &      1.00 &      0.95 &             $<$ 1.40 \\
 18 &   149.6073 &     2.6646 &      0.36 &     -0.60 &      1.22 &      0.82 &      0.38 &     -0.49 &      2.24 &      1.16 &             $<$ 1.26 \\
 19 &   149.9209 &     2.0312 &      0.36 &      2.36 &      1.84 &      1.25 &      0.41 &      1.28 &      1.26 &      0.96 &             $<$ 1.32 \\
 20 &   149.8058 &     2.2894 &      0.47 &     -0.06 &      1.57 &      0.99 &      0.43 &     -0.04 &      0.89 &      0.20 &             $<$ 1.92 \\
 21 &   150.1480 &     2.3108 &      0.48 &      3.48 &      1.60 &      1.23 &      0.47 &      2.17 &      1.12 &      0.10 &        5.30$\pm$2.44 \\
 22 &   150.5859 &     2.3711 &      0.43 &      1.43 &      0.99 &      0.77 &      0.30 &      1.45 &      1.07 &     -2.31 &             $<$ 1.20 \\
 23 &   150.1541 &     2.2262 &      0.35 &      0.66 &      1.16 &      1.09 &      0.44 &      0.57 &      0.92 &      0.53 &             $<$ 1.40 \\
 24 &   150.0049 &     2.5036 &      0.34 &      0.73 &      1.60 &      1.40 &      0.44 &      0.46 &      1.26 &      0.56 &             $<$ 1.35 \\
 25 &   150.5190 &     2.1952 &      0.37 &      0.30 &      1.09 &      1.06 &      0.28 &      3.81 &      0.71 &      0.91 &        0.97$\pm$0.26 \\
 26 &   150.0951 &     2.1937 &      0.42 &      0.58 &      1.59 &      1.15 &      0.41 &      0.36 &      1.12 &     -0.10 &             $<$ 1.62 \\
 27 &   150.4988 &     2.0749 &      0.42 &      0.45 &      1.26 &      0.74 &      0.28 &      0.35 &      1.11 &      0.81 &             $<$ 1.10 \\
 28 &   150.5009 &     2.6119 &      0.34 &      0.45 &      1.07 &      0.73 &      0.28 &      0.42 &      0.89 &      0.59 &             $<$ 0.85 \\
 29 &   150.0294 &     1.9970 &      0.39 &      0.35 &      1.62 &      1.61 &      0.41 &      3.94 &      0.63 &     -2.50 &        2.13$\pm$0.54 \\
 30 &   150.5183 &     2.3695 &      0.45 &     -0.86 &      1.00 &      0.72 &      0.27 &     -0.86 &      2.00 &      1.17 &             $<$ 1.15 \\
 31 &   150.3504 &     2.2755 &      0.47 &      1.75 &      1.15 &      0.65 &      0.28 &      1.51 &      1.55 &      1.18 &             $<$ 1.25 \\
 32 &   150.4906 &     2.5882 &      0.36 &      0.91 &      1.23 &      0.71 &      0.27 &      0.74 &      1.54 &     -0.13 &             $<$ 0.88 \\
 33 &   150.1707 &     2.5669 &      0.41 &     -0.87 &      1.08 &      0.98 &      0.41 &     -0.81 &      1.13 &     -2.61 &             $<$ 1.60 \\
 34 &   150.0071 &     1.7208 &      0.37 &      0.02 &      1.60 &      1.01 &      0.43 &      0.01 &      0.79 &     -1.70 &             $<$ 1.48 \\
 35 &   150.3371 &     2.2795 &      0.35 &      0.28 &      1.00 &      0.78 &      0.25 &      0.28 &      1.58 &      1.02 &             $<$ 0.81 \\
 36 &   150.0943 &     2.1393 &      0.22 &      1.06 &      1.63 &      1.67 &      0.44 &      3.84 &      0.67 &     -2.87 &        1.84$\pm$0.48 \\
 37 &   149.8424 &     2.3006 &      0.26 &      0.55 &      1.44 &      1.12 &      0.40 &      0.38 &      1.78 &     -2.65 &             $<$ 0.89 \\
  \\ 
\enddata\label{lowz} 
\tablenotetext{a}{The photometric redshifts and stellar masses of the galaxies are from \cite{ilb13} and the accuracy of those quantities is discussed in detail there. The SFRs are derived from the rest frame UV continuum and the infrared using Herschel PACS and SPIRE data as detailed in \cite{sco13}. All of the galaxies have greater than 10$\sigma$ photometry measurements so the uncertainties in M$_*$ and SFR associated with their measurements are always less than 10\%. As discussed in \cite{ilb13} the uncertainties in models used to 
derive the M$_*$ and SFR from the photometry are generally much larger but generally less than a factor 2.}
\tablenotetext{b}{SNR$_{tot}$ and SNR$_{peak}$ calculated separately from Eq. \ref{snr} and the column SNR lists the larger in absolute magnitude of those two SNRs. Note that we let the SNR be negative in cases where the flux estimate is negative so that several sigma negative flux values don't end up with a positive SNR above the detection thresholds.  }
\end{deluxetable*} 

\begin{deluxetable*}{rrrrrrrrrccr}
%\tabletypesize{0.2}
%\normaltsize
%\rotate
\tablecaption{\bf{Mid-z Galaxy Sample in Band 7} \label{tab1}  }
%\tablewidth{0pt}

\tablehead{ 
 \colhead{\#} &  \colhead{RA (2000)} & \colhead{Dec (2000)} & \colhead{z\tablenotemark{a} } &\colhead{S$_{tot}$} &\colhead{$\sigma_{tot}$} & \colhead{S$_{peak}$} &\colhead{$\sigma_{pix}$} & \colhead{SNR\tablenotemark{b}}  & \colhead{M$_{*}$\tablenotemark{a}} &  \colhead{Log(SFR)\tablenotemark{a}} &   \colhead{M$_{\rm ISM}$}  \\
 \colhead{} & \colhead{\deg} & \colhead{\deg} & \colhead{}  & \colhead{mJy} & \colhead{mJy}  & \colhead{mJy} & \colhead{mJy}  & \colhead{} &  \colhead{ $10^{11}$\msun} &  \colhead{\msun yr$^{-1}$}& \colhead{$10^{10}$\msun} }
\startdata 
& & & & & \bf{Mid-z} \\
  1 &   150.0728 &     1.7361 &      0.84 &      1.76 &      1.86 &      1.27 &      0.42 &      0.95 &      2.18 &     -0.67 &             $<$ 3.06 \\
  2 &   149.8392 &     2.2262 &      0.94 &      0.88 &      1.49 &      1.20 &      0.43 &      0.59 &      0.89 &     -0.60 &             $<$ 3.32 \\
  3 &   149.7302 &     2.0708 &      0.88 &      1.93 &      1.15 &      0.66 &      0.23 &      1.68 &      0.65 &     -1.04 &             $<$ 1.72 \\
  4 &   150.1133 &     2.0142 &      0.84 &     -0.12 &      0.90 &      0.56 &      0.21 &     -0.13 &      0.87 &     -3.00 &             $<$ 1.53 \\
  5 &   150.5422 &     2.2772 &      0.87 &     -1.61 &      1.03 &      0.32 &      0.21 &     -1.56 &      1.55 &     -0.78 &             $<$ 1.56 \\
  6 &   150.0771 &     2.5490 &      0.88 &     -2.12 &      1.46 &      1.17 &      0.47 &     -1.45 &      1.41 &     -0.40 &             $<$ 3.50 \\
  7 &   150.4510 &     2.4353 &      0.99 &     -0.38 &      0.66 &      0.59 &      0.21 &     -0.58 &      3.06 &     -1.11 &             $<$ 1.64 \\
  8 &   149.6546 &     2.7470 &      0.84 &     -0.01 &      0.72 &      0.49 &      0.16 &     -0.01 &      3.16 &     -3.00 &             $<$ 1.19 \\
  9 &   150.6991 &     2.3539 &      0.98 &      0.80 &      0.85 &      0.64 &      0.22 &      0.95 &      1.45 &     -3.00 &             $<$ 1.70 \\
 10 &   150.4478 &     1.8505 &      0.87 &      0.65 &      0.74 &      0.55 &      0.21 &      0.88 &      1.23 &     -3.00 &             $<$ 1.59 \\
 11 &   150.3820 &     2.1208 &      1.15 &     -0.23 &      0.83 &      0.50 &      0.20 &     -0.28 &      2.02 &     -3.00 &             $<$ 1.69 \\
 12 &   150.3798 &     2.0147 &      0.83 &      0.59 &      0.72 &      0.60 &      0.21 &      0.82 &      0.91 &      0.70 &             $<$ 1.53 \\
 13 &   149.8340 &     1.9810 &      0.94 &     -1.18 &      1.39 &      1.17 &      0.43 &     -0.85 &      2.00 &      0.40 &             $<$ 3.33 \\
 14 &   150.4944 &     2.2822 &      0.84 &      0.19 &      0.82 &      0.45 &      0.22 &      0.23 &      1.56 &     -3.00 &             $<$ 1.59 \\
 15 &   149.7526 &     2.2140 &      1.01 &      1.78 &      0.85 &      0.61 &      0.22 &      2.11 &      1.14 &      2.26 &        4.77$\pm$2.26 \\
 16 &   150.2639 &     1.9714 &      0.93 &     -0.57 &      0.65 &      0.46 &      0.20 &     -0.87 &      1.73 &      1.65 &             $<$ 1.57 \\
 17 &   149.9631 &     2.6131 &      0.90 &     -1.37 &      1.94 &      0.92 &      0.41 &     -0.71 &      0.89 &      0.90 &             $<$ 3.08 \\
 18 &   150.5358 &     2.2553 &      0.93 &      2.62 &      0.83 &      1.00 &      0.19 &      5.17 &      0.56 &      1.74 &        2.41$\pm$0.47 \\
 19 &   149.7205 &     2.7103 &      0.88 &      2.96 &      0.99 &      1.06 &      0.23 &      4.70 &      1.78 &      1.60 &        2.62$\pm$0.56 \\
 20 &   149.8618 &     2.1731 &      0.86 &      3.94 &      1.99 &      1.94 &      0.40 &      4.90 &      1.41 &      1.81 &        4.97$\pm$1.01 \\
 21 &   149.7728 &     1.8542 &      0.82 &     -0.50 &      1.63 &      0.79 &      0.41 &     -0.31 &      1.41 &      1.51 &             $<$ 2.96 \\
 22 &   149.8820 &     2.3506 &      0.86 &     -0.72 &      1.39 &      1.15 &      0.41 &     -0.52 &      0.88 &      1.42 &             $<$ 3.06 \\
 23 &   150.0685 &     2.2655 &      0.97 &      1.71 &      2.02 &      1.18 &      0.44 &      0.84 &      1.27 &      1.55 &             $<$ 3.47 \\
 24 &   149.9149 &     2.0944 &      0.85 &      0.06 &      1.88 &      1.22 &      0.43 &      0.03 &      1.14 &      0.21 &             $<$ 3.15 \\
 25 &   150.1289 &     2.1932 &      0.87 &      1.60 &      0.72 &      0.54 &      0.20 &      2.22 &      2.23 &      1.37 &        3.95$\pm$1.78 \\
 26 &   150.1089 &     2.0116 &      0.87 &      3.67 &      1.21 &      0.96 &      0.36 &      3.02 &      0.71 &      0.20 &        9.05$\pm$3.00 \\
 27 &   150.6456 &     1.9368 &      0.82 &     -0.18 &      0.76 &      0.50 &      0.19 &     -0.24 &      0.63 &      0.10 &             $<$ 1.37 \\
 28 &   149.7644 &     2.3401 &      0.81 &      1.04 &      0.93 &      0.69 &      0.23 &      1.11 &      1.78 &      1.92 &             $<$ 1.63 \\
 29 &   150.0906 &     2.5505 &      0.87 &     -0.01 &      1.43 &      0.61 &      0.33 &     -0.01 &      1.83 &     -0.11 &             $<$ 2.42 \\
 30 &   149.9190 &     2.6430 &      0.88 &      1.27 &      1.30 &      1.08 &      0.42 &      0.98 &      0.54 &     -0.67 &             $<$ 3.16 \\
 31 &   150.4074 &     1.8032 &      0.82 &      0.99 &      0.54 &      0.48 &      0.21 &      1.82 &      1.78 &      1.32 &             $<$ 1.51 \\
 32 &   149.9986 &     2.0634 &      0.91 &      1.76 &      0.84 &      1.39 &      0.39 &      2.09 &      1.26 &     -0.00 &        4.46$\pm$2.13 \\
 33 &   150.2983 &     2.0421 &      0.83 &      0.81 &      0.66 &      0.56 &      0.21 &      1.22 &      2.24 &     -0.50 &             $<$ 1.53 \\
  \\ 
\enddata\label{midz}
\tablenotetext{a}{The photometric redshifts and stellar masses of the galaxies are from \cite{ilb13} and the accuracy of those quantities is discussed in detail there. The SFRs are derived from the rest frame UV continuum and the infrared using Herschel PACS and SPIRE data as detailed in \cite{sco13}. All of the galaxies have greater than 10$\sigma$ photometry measurements so the uncertainties in M$_*$ and SFR associated with their measurements are always less than 10\%. As discussed in \cite{ilb13} the uncertainties in models used to 
derive the M$_*$ and SFR from the photometry are generally much larger but generally less than a factor 2.}
\tablenotetext{b}{SNR$_{tot}$ and SNR$_{peak}$ calculated separately from Eq. \ref{snr}  and the column SNR lists the larger in absolute magnitude of those two SNRs. Note that we let the SNR be negative in cases where the flux estimate is negative so that several sigma negative flux values don't end up with a positive SNR above the detection thresholds. }
\end{deluxetable*}

\begin{deluxetable*}{rrrrrrrrrccr}
%\tabletypesize{0.2}
%\normaltsize
%\rotate
\tablecaption{\bf{High-z Galaxy Sample in Band 7}}
\tablewidth{0pt}

\tablehead{ 
 \colhead{\#} &  \colhead{RA (2000)} & \colhead{Dec (2000)} & \colhead{z \tablenotemark{a}} &\colhead{S$_{tot}$} &\colhead{$\sigma_{tot}$} & \colhead{S$_{peak}$} &\colhead{$\sigma_{pix}$} & \colhead{SNR\tablenotemark{b}}  & \colhead{M$_{*}$\tablenotemark{a}} &  \colhead{Log(SFR)\tablenotemark{a}} &   \colhead{M$_{\rm ISM}$}  \\
 \colhead{} & \colhead{\deg} & \colhead{\deg} & \colhead{}  & \colhead{mJy} & \colhead{mJy}  & \colhead{mJy} & \colhead{mJy}  & \colhead{} &  \colhead{ $10^{11}$\msun} &  \colhead{\msun yr$^{-1}$}& \colhead{$10^{10}$\msun} }
\startdata 
& & & & & \bf{High-z} \\
  1 &   150.4938 &     1.7228 &      1.98 &      0.69 &      0.41 &      0.37 &      0.17 &      1.68 &      1.09 &      1.62 &             $<$ 1.64 \\
  2 &   150.3306 &     1.8136 &      1.46 &      0.05 &      0.59 &      0.71 &      0.31 &      0.09 &      1.26 &      2.17 &             $<$ 2.84 \\
  3 &   150.4812 &     1.9137 &      1.96 &     -0.07 &      0.44 &      0.46 &      0.18 &     -0.16 &      1.86 &      1.66 &             $<$ 1.74 \\
  4 &   149.9203 &     2.0204 &      1.95 &      3.19 &      0.37 &      3.10 &      0.19 &     16.70 &      1.57 &      2.19 &       10.13$\pm$0.61 \\
  5 &   150.3774 &     2.3118 &      1.97 &     -0.04 &      1.02 &      0.79 &      0.35 &     -0.04 &      2.20 &      2.29 &             $<$ 3.36 \\
  6 &   150.0107 &     2.3330 &      1.78 &      0.65 &      0.50 &      0.71 &      0.21 &      1.30 &      1.99 &      1.87 &             $<$ 1.96 \\
  7 &   149.7832 &     2.3716 &      2.23 &      1.85 &      0.44 &      1.33 &      0.20 &      6.70 &      1.05 &      2.40 &        4.42$\pm$0.66 \\
  8 &   149.5100 &     2.1023 &      1.97 &      1.19 &      0.46 &      1.51 &      0.17 &      8.70 &      0.92 &      2.36 &        4.89$\pm$0.56 \\
  9 &   149.8943 &     2.6216 &      1.51 &      0.27 &      0.63 &      0.43 &      0.19 &      0.43 &      0.36 &      1.31 &             $<$ 1.78 \\
 10 &   150.7518 &     2.2920 &      2.48 &      0.49 &      0.42 &      0.62 &      0.17 &      3.62 &      1.13 &      2.48 &        1.97$\pm$0.55 \\
 11 &   149.8151 &     2.8071 &      2.24 &      2.01 &      0.70 &      0.62 &      0.20 &      2.85 &      0.69 &      2.43 &        6.57$\pm$2.30 \\
 12 &   150.0250 &     2.1194 &      1.84 &      1.21 &      0.70 &      0.71 &      0.23 &      1.72 &      1.02 &      2.15 &             $<$ 2.24 \\
 13 &   150.3069 &     2.4549 &      2.31 &     -0.74 &      0.66 &      0.78 &      0.32 &     -1.13 &      0.63 &      0.69 &             $<$ 3.17 \\
 14 &   150.3298 &     2.4382 &      2.56 &      4.91 &      0.96 &      3.89 &      0.33 &     11.73 &      1.53 &      2.39 &       12.44$\pm$1.06 \\
 15 &   150.6614 &     2.0882 &      2.19 &      0.97 &      0.38 &      0.47 &      0.17 &      2.57 &      1.11 &      1.93 &        3.18$\pm$1.24 \\
 16 &   149.7104 &     2.5814 &      2.72 &     -0.02 &      0.57 &      0.48 &      0.18 &     -0.03 &      3.33 &      1.81 &             $<$ 1.82 \\
 17 &   150.3844 &     2.1799 &      1.53 &      0.32 &      0.85 &      1.31 &      0.30 &      4.39 &      0.81 &      1.22 &        4.24$\pm$0.97 \\
 18 &   150.1119 &     1.9774 &      2.54 &      0.62 &      0.52 &      0.53 &      0.22 &      1.19 &      0.95 &      1.51 &             $<$ 2.22 \\
 19 &   150.0233 &     2.1440 &      1.97 &      1.14 &      0.47 &      0.92 &      0.20 &      4.56 &      1.08 &      1.33 &        3.05$\pm$0.67 \\
 20 &   149.5147 &     2.4845 &      1.47 &      0.19 &      0.44 &      0.54 &      0.18 &      0.44 &      0.41 &      1.63 &             $<$ 1.62 \\
 21 &   149.6329 &     2.4728 &      2.13 &      0.93 &      0.48 &      0.68 &      0.17 &      4.02 &      1.03 &      1.36 &        2.18$\pm$0.54 \\
 22 &   150.2177 &     2.7733 &      1.94 &      4.16 &      1.01 &      2.73 &      0.31 &      8.67 &      2.17 &      2.42 &        8.28$\pm$0.95 \\
 23 &   150.3670 &     1.7233 &      2.10 &     -0.40 &      0.83 &      0.41 &      0.30 &     -0.48 &      0.86 &      1.78 &             $<$ 2.93 \\
 24 &   149.4926 &     2.8040 &      1.79 &      0.27 &      0.52 &      0.36 &      0.19 &      0.53 &      0.98 &     -0.04 &             $<$ 1.78 \\
 25 &   149.9330 &     1.8057 &      2.34 &     -0.10 &      0.48 &      0.51 &      0.20 &     -0.20 &      1.78 &      1.55 &             $<$ 1.93 \\
 26 &   149.6126 &     2.2638 &      2.32 &     -0.29 &      0.43 &      0.47 &      0.17 &     -0.66 &      0.61 &      1.87 &             $<$ 1.66 \\
 27 &   149.5978 &     1.7876 &      2.17 &      0.14 &      0.45 &      0.49 &      0.17 &      0.32 &      1.13 &      1.38 &             $<$ 1.66 \\
 28 &   150.1614 &     2.2350 &      1.91 &     -0.20 &      0.88 &      0.68 &      0.32 &     -0.23 &      1.51 &      1.46 &             $<$ 3.11 \\
 29 &   150.5190 &     2.5156 &      2.07 &      0.39 &      0.35 &      0.32 &      0.18 &      1.12 &      1.24 &      1.33 &             $<$ 1.71 \\
 30 &   150.0003 &     2.4312 &      2.05 &     -0.54 &      0.46 &      0.27 &      0.20 &     -1.18 &      1.74 &      2.05 &             $<$ 1.95 \\
 31 &   150.3792 &     2.5295 &      1.76 &     -0.07 &      0.72 &      0.70 &      0.31 &     -0.10 &      0.97 &      1.22 &             $<$ 2.95 \\
 \\
& & & & & \bf{IR bright} \\
 1 &   149.9094 &     1.9956 &      1.46 &      1.29 &      0.56 &      0.91 &      0.20 &      4.58 &      0.93 &      2.07 &        2.76$\pm$0.60 \\
  2 &   150.1153 &     2.0177 &      1.70 &      4.13 &      0.78 &      4.54 &      0.29 &     15.78 &      0.59 &      2.19 &       14.20$\pm$0.90 \\
  3 &   150.4655 &     2.4296 &      1.73 &      4.30 &      0.43 &      3.32 &      0.18 &     18.23 &      1.29 &      2.16 &       10.45$\pm$0.57 \\
  4 &   150.2149 &     2.5595 &      2.02 &      4.67 &      0.75 &      1.93 &      0.32 &      6.22 &      1.35 &      2.72 &       15.01$\pm$2.41 \\
  5 &   150.3936 &     1.9800 &      1.54 &      2.50 &      0.75 &      2.14 &      0.31 &      6.82 &      0.90 &      2.22 &        6.58$\pm$0.96 \\
  6 &   150.3367 &     2.1240 &      1.68 &      1.65 &      0.91 &      2.82 &      0.33 &      8.66 &      0.56 &      2.31 &        8.80$\pm$1.02 \\
  \\ 
\enddata\label{highz}
\tablenotetext{a}{The photometric redshifts and stellar masses of the galaxies are from \cite{ilb13} and the accuracy of those quantities is discussed in detail there. The SFRs are derived from the rest frame UV continuum and the infrared using Herschel PACS and SPIRE data as detailed in \cite{sco13}. All of the galaxies have greater than 10$\sigma$ photometry measurements so the uncertainties in M$_*$ and SFR associated with their measurements are always less than 10\%. As discussed in \cite{ilb13} the uncertainties in models used to 
derive the M$_*$ and SFR from the photometry are generally much larger but generally less than a factor 2.}
\tablenotetext{b}{SNR$_{tot}$ and SNR$_{peak}$ calculated separately from Eq. \ref{snr}  and the column SNR lists the larger in absolute magnitude of those two SNRs. Note that we let the SNR be negative in cases where the flux estimate is negative so that several sigma negative flux values don't end up with a positive SNR above the detection thresholds. }
\end{deluxetable*}

\end{document}